\documentclass[12pt]{iopart}

\newcommand{\be}{\begin{equation}}
\newcommand{\ee}{\end{equation}}
\newcommand{\ba}{\begin{eqnarray}}
\newcommand{\ea}{\end{eqnarray}}
\newcommand{\nn}{\nonumber}
\newcommand{\ep}{\epsilon}
\newcommand{\K}[1]{K \left(\frac{#1}{\Lambda}\right)}
\newcommand{\D}[1]{\Delta \left(\frac{#1}{\Lambda}\right)}
\newcommand{\Ld}[1]{\frac{\overrightarrow{\delta}}{\delta #1}}
\newcommand{\Rd}[1]{\frac{\overleftarrow{\delta}}{\delta #1}}
\newcommand{\lb}{\left\lbrace}
\newcommand{\rb}{\right\rbrace}
\newcommand{\SL}{S_\Lambda}
\newcommand{\SF}{S_{F,\Lambda}}
\newcommand{\SI}{S_{I,\Lambda}}
\newcommand{\fey}[1]{\hbox{$#1$\kern-0.5em\raise0.3ex\hbox{/}}}
\newcommand{\Op}{\mathcal{O}}
\newcommand{\N}{\mathcal{N}}
\newcommand{\vev}[1]{\left\langle #1 \right\rangle}
\newcommand{\comp}[1]{\left[ #1 \right]_\Lambda}
\begin{document}

\title[Composite operators in QED via ERG]{Gauge
  invariant composite operators of QED in the exact renormalization
  group formalism}

\author{H Sonoda}

\address{Physics Department, Kobe University, Kobe, 657-1547 Japan}

\ead{hsonoda@kobe-u.ac.jp}

\begin{abstract}
    Using the exact renormalization group (ERG) formalism, we study
    the gauge invariant composite operators in QED.  Gauge invariant
    composite operators are introduced as infinitesimal changes of the
    gauge invariant Wilson action.  We examine the dependence on the
    gauge fixing parameter of both the Wilson action and gauge
    invariant composite operators.  After defining ``gauge fixing
    parameter independence,'' we show that any gauge independent
    composite operators can be made ``gauge fixing parameter
    independent'' by appropriate normalization.  As an application, we
    give a concise but careful proof of the Adler-Bardeen
    non-renormalization theorem for the axial anomaly in an arbitrary
    covariant gauge by extending the original proof by A.~Zee.
\end{abstract}

\pacs{11.10.Gh, 11.15.-q, 11.15.Bt}


\maketitle

\section{Introduction}
\label{introduction}

In QED, as in any gauge theories, all physical quantities are gauge
invariant.  They are often given as gauge invariant composite
operators such as the electric current or energy-momentum tensor.  The
purpose of this paper is to study gauge invariant composite operators
in QED formulated with the exact renormalization group
(ERG).\footnote[1]{There are many reviews of ERG.
  \cite{Igarashi:2009tj}, which gives references to some of the
  earlier reviews, has an emphasis on perturbative applications of
  ERG.  A most recent review is \cite{Rosten:2010vm}.}  We are
especially interested in the dependence of the gauge invariant
composite operators on the covariant gauge fixing parameter.  We will
see that particular normalization convention must be adopted for the
independence of the composite operators on the gauge fixing parameter.

There are many formulations for QED (and YM theories), and accordingly
gauge invariant composite operators have already been studied in
various formulations.  Most notably, gauge invariant composite
operators in YM theories have been studied with the dimensional
regularization.  (See Chapters 12 \& 13 of \cite{Collins} and
references therein.)  Within the ERG formalism, however, gauge
invariant composite operators have not been fully studied, and we wish
to fill the gap in this paper.

The advantage of formulating QED with ERG is three-fold: first, the
ease of renormalization.  The Wilson action $\SL$ \cite{Wilson:1973jj}
with a finite UV cutoff $\Lambda$ is obtained as a perturbative
solution to the ERG differential equation.\cite{Polchinski:1983gv}
There is no need for regularization or taking a limit for
renormalization.  Renormalization is done with a selection of
solutions with appropriate behaviors for large $\Lambda$.  Second, the
gauge invariance is incorporated nicely as the Ward-Takahashi (WT)
identity among composite operators, even though the gauge invariance
is not manifest in the Wilson action.  Third, we work with a fixed
number of dimensions for space(time), and the chirality of fermion
fields can be introduced with ease.  Hence, this formalism is suited
for the study of axial or chiral anomalies.

In any perturbative formulation of QED, including ERG, gauge
invariance is incorporated as the WT identities for the renormalized
correlation functions.\footnote{A notable exception is the manifestly
  gauge invariant ERG formalism of Arnone, Morris, and Rosten.  (See
  \cite{Morris:2006in} and references therein.  See also IX B of
  \cite{Rosten:2010vm}.)}  Within the ERG formalism alone, there are
several ways.\cite{Becchi:1996an, Ellwanger:1994iz, Bonini:1993kt,
  Reuter:1994sg} The discussion in this paper is based on a
formulation developed in \cite{Sonoda:2007dj} (which is based closely
on the earlier work of Becchi\cite{Becchi:1996an}).  The main tool is
composite operators which are defined as infinitesimal variations of
the Wilson action.  Since the Wilson action and composite operators
are well-defined functionals of field variables, formal manipulations
acting on these functionals, such as functional differentiation, are
also well defined.  We have no regularization to take to a limit; we
need not worry if the operator equations remain valid in the limit.

The main subject of this paper is the dependence of the Wilson action
and composite operators on the covariant gauge fixing parameter,
denoted as $\xi$.  Introduction of $\xi$ is only for the convenience
of perturbative expansions, and physics should not depend on the
arbitrary choice of $\xi$.  Our first task is to translate this simple
requirement into an equation satisfied by gauge invariant composite
operators.  After defining ``$\xi$-independence'' of composite
operators, we will show that any gauge invariant composite operators
can be made ``$\xi$-independent'' by appropriate normalization.  We
will then study the anomalous dimensions of gauge invariant \&
$\xi$-independent composite operators, and show them to be independent
of $\xi$.  Given an anomalous dimension as a function of the squared
gauge coupling $e^2$, a composite operator is determined uniquely
except for a constant factor.

We apply our results to the Adler-Bardeen non-renormalization theorem
for the axial anomaly.\cite{Adler:1969er} We follow the proof given by
A.~Zee who applied the renormalization group for the first time to the
non-renormalization of the axial anomaly.\cite{Zee:1972zt} Our
treatment resembles Chapter 13 of \cite{Collins}, where the
dimensional regularization is used instead of ERG.  The apparent
resemblance is inevitable; the main difference is in the technique of
constructing gauge invariant composite operators.  We pay a careful
attention to the gauge fixing parameter, necessary for the
perturbative formulation of QED.  Since the anomaly equation is
formulated as a linear relation among gauge invariant \&
$\xi$-independent composite operators, our proof is valid in any
gauge.

The organization of the paper is as follows.  In Section 2 we review
the ERG formalism for QED that gives the Ward-Takahashi (WT)
identities in terms of an operator identity. For the reader unfamiliar
with perturbative ERG, we have prepared a short summary of the
formalism in Appendix A, using the example of the $\phi^4$ theory.
The formalism is simple both logically and technically, and we hope
that Section 2 and Appendix A are more than enough to prepare the
reader to follow the remaining sections.  In Section 3 we introduce a
WT identity for composite operators to define their gauge invariance.
In Sections 4 and 5, we discuss the main subject of this paper, the
$\xi$-dependence.  In Section 4, we discuss the $\xi$-dependence of
the Wilson action, and derive an explicit formula for the
$\xi$-dependence of renormalized correlation functions of elementary
fields.  This extends the formula originally given by Landau and
Khalatnikov for the bare correlation functions.\cite{Landau:1955zz,
  Svidzinskii, Bogoliubov, Zumino:1959wt} In Section 5, we extend the
discussion to composite operators.  We first define what we mean by
$\xi$-independent composite operators, and then show that any gauge
invariant composite operator can be made $\xi$-independent by
normalization.  In Section 6 we relate ERG to the standard RG by
discussing the $\mu$-dependence of the Wilson
action.\cite{Sonoda:2006ai} A renormalization scale $\mu$ is
introduced to specify a Wilson action as a unique solution to the ERG
differential equation.  We introduce the beta functions of the
parameters and anomalous dimensions of the elementary fields.  Then,
in Section 7, we define the anomalous dimensions of composite
operators.  We show that the anomalous dimension of a gauge invariant
\& $\xi$-independent composite operator is independent of $\xi$.  In
Section 8, we apply the results to the non-renormalization of the
axial anomaly.

Most of the technicalities have been relegated to the appendices which
are almost as long as the main text.  We wish to avoid interrupting
the simple logical flow of the paper by technical details, even though
the details of how to use the techniques of ERG constitute an essential
part of this work.  The appendices give all the necessary details
except the actual 1-loop calculations of which we only enumerate the
results.

Throughout the paper we work with the four dimensional Euclidean
space.  We find the following shorthand notations convenient and use
them frequently:
\be
\int_p \equiv \int \frac{d^4 p}{(2 \pi)^4},\quad
\delta (p) \equiv (2 \pi)^4 \delta^{(4)} (p)
\ee

\section{The Wilson action for QED}
\label{Wilson}

In the exact renormalization group formalism, the main object of our
study is a Wilson action.\cite{Wilson:1973jj} In the case of QED, a
Wilson action $\SL [A_\mu, \psi, \bar{\psi}]$ is a functional of the
photon field $A_\mu$ and Dirac spinor fields $\psi, \bar{\psi}$ with a
finite UV cutoff $\Lambda$.\footnote{For the reader unfamiliar with
  the (perturbative) ERG formalism, we have prepared a short summary
  in \ref{nutshell}.}  We define the correlation functions of
elementary fields by
\be \vev{\cdots}_{\SL} \equiv \int [dA_\mu d\psi
d\bar{\psi}]\,\e^{\SL} \,\cdots\Big/ \int [dA_\mu d\psi d\bar{\psi}]
\,\e^{\SL} 
\ee 
where the dots denote a product of elementary fields.  Given the free
action 
\ba \SF &\equiv& - \int_k \frac{1}{\K{k}} \left(k^2
    \delta_{\mu\nu} - \left(1 - \frac{1}{\xi}\right) k_\mu k_\nu
\right) \frac{1}{2} A_\mu
(k) A_\nu (-k)\\
&& - \int_p \frac{1}{\K{p}} \bar{\psi} (-p) \left( \fey{p} + i m
\right) \psi (p) 
\ea
where $\xi$ is introduced as a gauge fixing parameter, the propagators
are obtained as
\ba 
\vev{A_\mu (k)
  A_\nu (k')}_{\SF} &=& \frac{\K{k}}{k^2} \left( \delta_{\mu\nu} -
    (1-\xi) \frac{k_\mu k_\nu}{k^2} \right) \delta
(k+k')\\
\vev{\psi (p) \bar{\psi} (-q)}_{\SF} &=& \frac{\K{p}}{\fey{p} + i m}
\delta (p-q) 
\ea
The cutoff function $\K{k}$ is a decreasing but positive function of
$k^2/\Lambda^2$: it is $1$ for $k^2 < \Lambda^2$, and is nowhere zero
except at infinity so that the division by $\K{k}$ makes sense for
finite $k^2$.  We also assume that it decays fast enough as $k^2 \to
\infty$ so that the functional integrals are well defined, free of UV
divergences.\footnote{$K(k) = \mathrm{o} \left(k^{-4}\right)$ for $k^2
  \gg 1$ is sufficient.}  The cutoff dependence of the Wilson action
is given by the ERG differential equation\cite{Polchinski:1983gv}
\ba \label{ERG-qed}&&-
\Lambda \partial_\Lambda \e^{\SL} = \int_k \D{k} \left[
    \frac{1}{\K{k}} A_\mu (k) \frac{\delta}{\delta A_\mu
      (k)}\right.\nn\\
&& \qquad\qquad + \left.  \frac{1}{k^2} \left( \delta_{\mu\nu} - (1-\xi)
        \frac{k_\mu k_\nu}{k^2} \right) \frac{1}{2}
    \frac{\delta^2}{\delta A_\mu
      (k) \delta A_\nu (-k)} \right] \e^{\SL}\nn\\
&& \quad+ \int_p \D{p} \left[ \frac{1}{\K{p}} \lb \e^{\SL} \Rd{\psi
      (p)} \psi (p) + \bar{\psi}
    (-p) \Ld{\bar{\psi} (-p)} \e^{\SL}\rb \right.\nn\\
&&\left.\qquad\qquad + \Tr \Ld{\bar{\psi} (-p)} \e^{\SL} \Rd{\psi (p)}
    \frac{1}{\fey{p} + i m} \right] \ea 
where we denote
\be
\D{k} \equiv \Lambda \frac{\partial}{\partial \Lambda} \K{k}
\ee
(\ref{ERG-qed}) gives the cutoff dependence of the Wilson action, but
at the same time it assures that the following linear combination of
correlation functions remains independent of $\Lambda$
\cite{Sonoda:2007dj} (see also ``dual actions'' in
\cite{Rosten:2010vm}):
\ba
&&\vev{A_{\mu_1} (k_1) \cdots A_{\mu_L} (k_L) \psi (p_1) \cdots \psi
  (p_N) \bar{\psi} (-q_1) \cdots \bar{\psi} (-q_N)}^\infty\nn\\
&\equiv& \prod_{i=1}^L \frac{1}{\K{k_i}} \prod_{j=1}^N
\frac{1}{\K{p_j} \K{q_j}} \Big[\, \vev{A_{\mu_1} (k_1)
      \cdots}_{\SL}\nn\\
&& + \sum_{i<j} \frac{\K{k_i} \left(\K{k_i}-1\right)}{k_i^2} \left(
    \delta_{\mu_i \mu_j} - (1-\xi) \frac{k_{i, \mu_i} k_{i,
        \mu_j}}{k_i^2}\right) \nn\\
&&\qquad\qquad \times \delta (k_i+k_j) \vev{\cdots
  \widehat{A_{\mu_i} (k_i)} \cdots \widehat{A_{\mu_j} (k_j)} \cdots}_{\SL} \nn\\
&& + \sum_{i,j} \frac{\K{p_i} \left(\K{p_i} - 1\right)}{\fey{p}_i + i
  m} \delta (p_i-q_j) \vev{\cdots \widehat{\psi (p_i)} \cdots
  \widehat{\bar{\psi} (-q_j)} \cdots}_{\SL} \nn\\
&& + \cdots \quad\Big]\label{corr-infty}
\ea
where the dots indicate replacement of more pairs.  Especially, the
cutoff independent two-point functions are obtained as
\ba
\vev{A_\mu (k) A_\nu (k')}^\infty &\equiv& \frac{1}{\K{k}^2}
\vev{A_\mu (k) A_\nu (k')}_{\SL} \nn\\
&& + \frac{1 - 1/\K{k}}{k^2} \left( \delta_{\mu\nu} - (1-\xi)
    \frac{k_\mu k_\nu}{k^2} \right) \delta (k+k')\\
\vev{\psi (p) \bar{\psi} (-q)}^\infty &\equiv& \frac{1}{\K{p}^2}
\vev{\psi (p) \bar{\psi} (-q)}_{\SL} + \frac{1-1/\K{p}}{\fey{p} + i m}
\delta (p-q)
\ea

To specify a solution of the ERG differential equation uniquely, we
can impose an asymptotic condition on the Wilson action for large
$\Lambda$.  Perturbative renormalizability of QED amounts to the
existence of Wilson actions for which the coefficients of higher
dimensional terms vanish as $\Lambda \to \infty$.\cite{Sonoda:2002pb}
More specifically, for the field momenta small compared with
$\Lambda$, we can expand the action as
\ba \SL &\stackrel{\Lambda \to \infty}{\longrightarrow}& - \int d^4
x\, \left[ \frac{1}{2} \partial_\mu A_\nu \partial_\mu A_\nu - \left(1
        - \frac{1}{\xi}\right) \frac{1}{2} \left(\partial_\mu
        A_\mu\right)^2 + \bar{\psi} \left( \frac{1}{i} \fey{\partial}
        + i m \right) \psi
\right]\nn \\
&& + \int d^4 x\, \left[ a_2 (\Lambda) \frac{1}{2} A_\mu^2 + z (\Lambda)
    \frac{1}{2} (\partial_\mu A_\nu)^2 + \tilde{z} (\Lambda)
    \frac{1}{2} (\partial \cdot A)^2\right.\\
&& \quad \left.+ \bar{\psi} \left( z_F (\Lambda) \frac{1}{i}
        \fey{\partial} + z_m (\Lambda) i m \right) \psi + a_3
    (\Lambda) \bar{\psi} \fey{A} \psi  + 
a_4 (\Lambda) \frac{1}{8} \left(A_\mu^2\right)^2 \right]\nn
\ea
where the first integral comes from $\SF$, and the rest from the
interaction part 
\be
\SI \equiv \SL - \SF
\ee
The cutoff dependence of the coefficients are determined by
(\ref{ERG-qed}), but their values at a finite momentum scale $\mu$ can
be chosen arbitrarily.  Hence, the Wilson action is parametrized by
seven arbitrary parameters:
\[
a_2 (\mu),\quad z(\mu),\quad \tilde{z} (\mu),\quad z_F (\mu),\quad z_m
(\mu),\quad a_3 (\mu),\quad a_4 (\mu)
\]
All are dimensionless, except for $a_2 (\mu)$ which has mass dimension
$2$.  The Ward-Takahashi identity, reviewed in the next section,
reduces the number of arbitrary parameters from seven to
three.\cite{Sonoda:2007dj}

\section{Gauge invariant composite operators}
\label{comp}

A composite operator $\Op_\Lambda [A_\mu, \psi, \bar{\psi}]$ is a
functional which can be regarded as an infinitesimal change of the
Wilson action.\cite{Becchi:1996an, Igarashi:2009tj} Hence,
$\Op_\Lambda \e^{\SL}$ satisfies the same ERG differential equation as
$\e^{\SL}$:
\ba \label{ERG-qed-comp}
&& - \Lambda \partial_\Lambda \left( \Op_\Lambda \e^{\SL}\right) =
\int_k \D{k} \left[ \frac{1}{\K{k}} A_\mu (k) \frac{\delta}{\delta
      A_\mu
      (k)}\right.\nn\\
&& \qquad\qquad + \left.  \frac{1}{k^2} \left( \delta_{\mu\nu} - (1-\xi)
        \frac{k_\mu k_\nu}{k^2} \right) \frac{1}{2}
    \frac{\delta^2}{\delta A_\mu
      (k) \delta A_\nu (-k)} \right] \left(\Op_\Lambda \e^{\SL}\right)\nn\\
&& \quad+ \int_p \D{p} \left[ \frac{1}{\K{p}} \lb \left(\Op_\Lambda
        \e^{\SL}\right)  \Rd{\psi (p)} \psi (p) + \bar{\psi}
    (-p) \Ld{\bar{\psi} (-p)} \left(\Op_\Lambda \e^{\SL}\right)\rb \right.\nn\\
&&\left.\qquad\qquad + \Tr \Ld{\bar{\psi} (-p)} \left(\Op_\Lambda
        \e^{\SL}\right) \Rd{\psi (p)} \frac{1}{\fey{p} + i m} \right]
\ea
This is obtained from (\ref{ERG-qed}) by replacing $\e^{\SL}$ by
$\Op_\Lambda \e^{\SL}$.  The above equation implies that the
correlation functions
\ba
&&\vev{\Op\,A_{\mu_1} (k_1) \cdots A_{\mu_N} (k_L) \psi (p_1) \cdots \psi
  (p_N) \bar{\psi} (-q_1) \cdots \bar{\psi} (-q_N)}^\infty\nn\\
&\equiv& \prod_{i=1}^L \frac{1}{\K{k_i}} \prod_{j=1}^N
\frac{1}{\K{p_j} \K{q_j}} \Big[ \,\vev{\Op_\Lambda\, A_{\mu_1} (k_1)
      \cdots}_{\SL}\nn\\
&& + \sum_{i<j} \frac{\K{k_i} \left(\K{k_i}-1\right)}{k_i^2} \left(
    \delta_{\mu_i \mu_j} - (1-\xi) \frac{k_{i, \mu_i} k_{i,
        \mu_j}}{k_i^2}\right) \nn\\
&&\qquad\qquad \times \delta (k_i+k_j) \vev{\Op_\Lambda\,\cdots
  \widehat{A_{\mu_i}(k_i)} \cdots \widehat{A_{\mu_j}(k_j)} \cdots}_{\SL} \nn\\
&& + \sum_{i,j} \frac{\K{p_i} \left(\K{p_i} - 1\right)}{\fey{p}_i + i
  m} \delta (p_i-q_j) \vev{\Op_\Lambda\,\cdots \widehat{\psi (p_i)} \cdots
  \widehat{\bar{\psi} (-q_j)} \cdots}_{\SL} \nn\\
&& + \cdots \quad\Big]\label{Op-corr-infty}
\ea
are independent of the cutoff $\Lambda$.  

The simplest examples of composite operators are given by
\ba
\left[A_\mu (k)\right]_\Lambda &\equiv& \frac{1}{\K{k}} A_\mu (k)
\nn\\
&&\quad +
\frac{1-\K{k}}{k^2} \left(\delta_{\mu\nu} - (1-\xi) \frac{k_\mu
      k_\nu}{k^2} \right) \frac{\delta \SL}{\delta A_\nu (-k)}\\
\left[\psi (p)\right]_\Lambda &\equiv& \frac{1}{\K{p}} \psi (p) + \frac{1 -
  \K{p}}{\not{p} + i m} \Ld{\bar{\psi} (-p)} \SL\\
\left[\bar{\psi} (-p)\right]_\Lambda &\equiv& \frac{1}{\K{p}}
\bar{\psi} (-p) + \SL \Rd{\psi (p)} \frac{1 - \K{p}}{\not{p} + i m}
\ea
These correspond to the respective elementary fields in the sense that
\be
\lb
\begin{array}{r@{~=~}l}
\vev{[A_\mu (k)] \cdots}^\infty & \vev{A_\mu (k) \cdots}^\infty\\
\vev{[\psi (p)] \cdots}^\infty & \vev{\psi (p) \cdots}^\infty\\
\vev{[\bar{\psi} (-p)] \cdots}^\infty & \vev{\bar{\psi} (-p)
  \cdots}^\infty
\end{array}\right.
\ee
where the dots represent the same product of elementary fields on both
sides; the left-hand sides are defined by (\ref{Op-corr-infty}) while
the right-hand sides are defined by (\ref{corr-infty}).

In constructing QED, the gauge invariance plays the most important
role.  The gauge invariance of QED is realized as the Ward-Takahashi
(WT) identities among the correlation functions.  In
\cite{Sonoda:2007dj} the WT identity of the Wilson action has been
given concisely as the current conservation equation:
\be\label{oldWT}
k_\mu J_\mu (k) = e \Phi (k)
\ee
where the two composite operators are defined by\footnote{In
  \cite{Sonoda:2007dj}, $- e \Phi (k)$ is denoted as $\Phi (k)$.}
\ba
J_\mu (k) &\equiv& \frac{\delta \SI}{\delta A_\mu (-k)}\\
\Phi (k) &\equiv& \int_p \K{p} \e^{-\SL} \left[ - \Tr \left( \e^{\SL} \comp{\psi
        (p+k)} \right) \Rd{\psi (p)}\right.\nn\\
&&\qquad \left. + \Tr \Ld{\bar{\psi} (-p)} \left(
        \e^{\SL} \comp{\bar{\psi} (-p+k)} \right) \right]\label{Phi}
\ea
$J_\mu (k)$ defines the electric current.  $\Phi (k)$ is an
``equation-of-motion'' composite operator whose correlation functions
are given exactly by\footnote{An ``equation-of-motion'' composite
  operator $\Op_\Lambda$ has the property that $\Op_\Lambda
  \,\e^{\SL}$ is a total derivative with respect to fields.}
\ba
&&\vev{\Phi (k)\, A_{\mu_1} (k_1) \cdots \psi (p_1) \cdots \bar{\psi}
  (-q_1) \cdots}^\infty\\
&& = \sum_i \lb - \vev{A_{\mu_1} (k_1)\cdots \psi (p_i + k) \cdots}^\infty
+ \vev{A_{\mu_1} (k_1) \cdots \bar{\psi} (-q_i + k) \cdots}^\infty \rb\nn
\ea
where each $\psi (p_i)$ is replaced by $- \psi (p_i+k)$, and each
$\bar{\psi} (-q_i)$ by $+ \bar{\psi} (-q_i+k)$.  

We can make the implication of (\ref{oldWT}) more transparent by
rewriting it as
\be\label{WT-qed}
k^2 \frac{1}{\xi} k_\mu \left[ A_\mu (k) \right] = D(k) + e \Phi (k)
\ee
where 
\be\label{D}
D(k) \equiv - \K{k} k_\mu
\frac{\delta \SL}{\delta A_\mu (-k)}
\ee
is also an equation-of-motion composite operator just like $\Phi (k)$;
it eliminates photon fields one by one:
\ba
&&\vev{D(k)\,A_{\mu_1}
  (k_1) \cdots A_{\mu_L} (k_L) \psi (p_1) \cdots }^\infty\nn\\
&&= \sum_{i=1}^L \delta (k+k_i) k_{\mu_i} \vev{A_{\mu_1} (k_1) \cdots
  \widehat{A_{\mu_i} (k_i)} \cdots A_{\mu_L} (k_L) \psi (p_1) \cdots
}^\infty \ea 
For the correlation functions of elementary fields, (\ref{WT-qed})
gives immediately
\ba\label{WT-qed-corr}
&&\frac{1}{\xi} k_\mu \vev{A_\mu (k) \cdots}^\infty\nn\\
&& = \frac{1}{k^2} \Big[
\sum_i k_{\mu_i} \delta (k+k_i) \vev{\cdots\widehat{A_{\mu_i}
    (k_i)}\cdots}^\infty\\
&&\quad + e \sum_i \lb - \vev{\cdots \psi (p_i+k)\cdots}^\infty
+ \vev{\cdots\bar{\psi} (-q_i+k)\cdots}^\infty \rb\Big]\nn
\ea
which is the usual form of the WT identities in QED.

As a consequence of (\ref{WT-qed}) (or equivalently (\ref{oldWT})),
the number of arbitrary parameters is reduced from seven to three, as
has been discussed in \cite{Sonoda:2007dj}. We can take the three
parameters as
\[
z(\mu),\quad z_F (\mu),\quad z_m (\mu)
\]
corresponding to the freedom of normalizing $A_\mu$, $\psi$ \&
$\bar{\psi}$, and the mass parameter $m$.  The parameter $e$ is
introduced via the WT identity (\ref{WT-qed}).  In general we can
choose the above three parameters as arbitrary functions of $e^2$ and
$\xi$ (the gauge fixing parameter).  In \cite{Sonoda:2007dj} we have
chosen these three parameters as zero.  Though this choice is
practical, it is not a good choice if we wish to control the
dependence of $\SL$ on $\xi$.  This will be explained in the next
section.

Before proceeding to the next section, let us generalize the WT
identity for the composite operators to define their gauge invariance.
We introduce the WT identity of a composite operator $\Op_\Lambda$ by
\be\label{WT-qed-op}
\frac{1}{\xi} k_\mu \comp{ A_\mu (k) \Op}
= \frac{1}{k^2} \left( D(k) + e \Phi (k) \right) * \Op_\Lambda
\ee
The left-hand side is a composite operator corresponding to the
product of $A_\mu (k)$ with $\Op_\Lambda$:
\ba
\comp{A_\mu (k) \Op}
&\equiv& \comp{ A_\mu (k) } \Op_\Lambda +
\frac{1-\K{k}}{k^2} \left( \delta_{\mu\nu} - (1-\xi) \frac{k_\mu
      k_\nu}{k^2} \right) \frac{\delta \Op_\Lambda}{\delta A_\nu
  (-k)}\nn\\
&=& \e^{- \SL} \left[ \frac{1}{\K{k}} A_\mu (k) + \frac{1-\K{k}}{k^2} \right.\\
&& \qquad\qquad\left.\times \left(\delta_{\mu\nu} - (1-\xi) \frac{k_\mu
      k_\nu}{k^2} \right) \frac{\delta}{\delta A_\nu (-k)} \right]
\left( \e^{\SL} \Op \right)\nn
\ea
To obtain this, we consider $\comp{A_\mu (k)} \e^{\SL}$, vary $\SL$
infinitesimally by $\Op_\Lambda$, and divide the result by $\e^{\SL}$.
(See \ref{app-phi4-comp} for the product of a composite operator with
an elementary field.)  The product is defined so that
\be
\vev{\left[A_\mu (k) \Op\right] \cdots}^\infty = \vev{A_\mu (k) \Op
  \cdots}^\infty
\ee
The star products on the right-hand side of (\ref{WT-qed-op}) are
defined as the following equations-of-motion operators\footnote{Note
  $D(k)$ of (\ref{D}) is the same as $D(k)*1$.  Similarly, $\Phi (k) =
\Phi (k) * 1$.}:
\ba
D(k) * \Op_\Lambda &\equiv& \e^{-\SL} (-) \frac{\delta}{\delta
  A_\mu (-k)} \left( k_\mu \Op_\Lambda \e^{\SL} \right)\label{Dstar}\\
\Phi (k) * \Op_\Lambda &\equiv& \e^{-\SL} \int_p \K{p}
\lb - \Tr \left( \comp{\Op\,\psi (p+k)} \e^{\SL} \right)
\Rd{\psi (p)} \right.\nn\\
&& \left.\qquad + \Tr \Ld{\bar{\psi} (-p)} \left( \comp{\Op\,\bar{\psi}
        (-p+k)} \e^{\SL} \right) \rb\label{Phistar}
\ea
As we obtain $\comp{A_\mu (k) \Op} \e^{\SL}$ from $\comp{A_\mu (k)}
\e^{\SL}$, we obtain $\e^{\SL} D(k) * \Op_\Lambda$ and $\e^{\SL} \Phi
(k)* \Op_\Lambda$ from $\e^{\SL} D(k)$ and $\e^{\SL} \Phi (k)$ by
changing $\SL$ infinitesimally by $\Op_\Lambda$.  The correlation
functions of these composite operators are given by
\ba
\vev{D(k)*\Op\, \cdots }^\infty &=&
\sum_i \delta (k+k_i) k_{\mu_i} \vev{\Op\, \cdots \widehat{A_{\mu_i}
    (k_i)} \cdots}^\infty\\
\vev{\Phi (k)*\Op\, \cdots }^\infty &=& \sum_i
\lb - \vev{\Op\,\cdots\psi (p_i+k)\cdots}^\infty \right.\nn\\
&&\qquad \left.+ \vev{\Op\, \cdots
  \bar{\psi} (-q_i + k)\cdots}^\infty\rb
\ea

We have thus obtained the WT identity (\ref{WT-qed-op}) for composite
operators from the WT identity (\ref{WT-qed}) for the Wilson action by
varying $\SL$ infinitesimally by $\Op_\Lambda$.  Before concluding
this section, we give the implication of (\ref{WT-qed-op}) for the
correlation functions:
\ba\label{WT-qed-op-corr}
&& \frac{1}{\xi} k_\mu \vev{A_\mu (k) \Op\, \cdots}^\infty\nn\\
&&= \frac{1}{k^2} \Big[ \sum_i k_{\mu_i} \delta (k+k_i) \, \vev{\Op\, \cdots
  \widehat{A_{\mu_i} (k_i)} \cdots}^\infty\\
&&\quad+ e \sum_i \lb - \vev{\Op\, \cdots \psi (p_i + k)\cdots}^\infty
+ \vev{\Op\, \cdots \bar{\psi} (-q_i + k)\cdots}^\infty \rb \Big]\nn
\ea

\section{Dependence of the action on the gauge fixing parameter $\xi$} 
\label{xi}

We are now ready to discuss the main subject of this paper: how the
Wilson action and gauge invariant composite operators depend on the
gauge fixing parameter $\xi$.  Of course we do not expect any physical
quantities to depend on $\xi$, but we cannot remove $\xi$-dependence
entirely from either the Wilson action or the gauge invariant
composite operators.  Fortunately, for QED, we can derive the
$\xi$-dependence of the correlation functions explicitly.

In this section we consider only the Wilson action and derive the
renormalized Landau-Khalatnikov relation that gives the $\xi$
dependence of the correlation functions.\cite{Landau:1955zz,
  Svidzinskii, Bogoliubov, Zumino:1959wt} In order
to present the logical flow clearly, we defer technical details to
\ref{appendix-equality}.

We first introduce a composite operator $\Op_\xi$ that generates an
infinitesimal variation of $\xi$:
\be\label{Oxi}
\Op_\xi \equiv - \e^{-\SL} \left[ \partial_\xi + \int_k
    \frac{\K{k}\left(\K{k}-1\right)}{k^4} k_\mu k_\nu \frac{1}{2}
    \frac{\delta^2}{\delta A_\mu (k) \delta A_\nu (-k)} \right]
\e^{\SL}
\ee
We may call $\Op_\xi$ a composite operator ``conjugate to'' $\xi$.  It
has the following correlation functions:
\be\label{Oxi-corr}
\vev{\Op_\xi \, \cdots}^\infty
= - \frac{\partial}{\partial \xi} \vev{\cdots}^\infty
\ee
where the dots represent a product of elementary fields.  The second
term of (\ref{Oxi}) arises due to the $\xi$ dependence of the photon
propagator.

Let us recall that our Wilson action depends not only on $m, e, \xi$,
but also on $z_m (\mu), z (\mu), z_F (\mu)$.  In perturbation theory
we can expand the latter three in powers of $e^2$, but the
coefficients of the expansions can be given arbitrary $\xi$ dependence.
We would like to choose 
\[
\partial_\xi z_m (\mu), \quad \partial_\xi z (\mu),\quad \partial_\xi
z_F (\mu)
\]
in such a way that $\Op_\xi$ becomes an equation-of-motion composite
operator.  Without this choice, $\Op_\xi$ would mix with the composite
operators conjugate to $e$ and $m$ ($\Op_e$ and $\Op_m$, to be defined
in section \ref{RG for QED}) so that $\Op_\xi$ would not be an
equation-of-motion composite operator.  This convenient choice will
give us immediately a renormalized Landau-Khalatnikov relation for the
correlation functions of the elementary fields.

To motivate the desired rewriting of $\Op_\xi$, let us first compute
it classically.  Since the classical action is
\be
\hspace{-1cm}S_{\mathrm{cl}} \equiv - \int d^4 x\, \left[
\frac{1}{2} \left(\partial_\mu A_\nu\right)^2 - \left(1 -
    \frac{1}{\xi}\right) \frac{1}{2} \left(\partial_\mu A_\mu\right)^2
+ \bar{\psi} \left( \frac{1}{i} \fey{\partial} + i
        m - e \fey{A} \right) \psi \right]
\ee
we obtain
\be\label{Oxi-classical} \left(\Op_\xi\right)_{\mathrm{cl}} =
- \partial_\xi S_{\mathrm{cl}} = - \frac{1}{\xi^2} \int d^4 x\,
\frac{1}{2} \left(\partial_\mu A_\mu\right)^2 \ee
We would like to find a composite operator that corresponds to this
classical expression. 

We first try
\be
- \frac{1}{\xi^2} \int_k \frac{1}{2} k_\mu k_\nu \comp{ A_\mu (k)
  A_\nu (-k)}
\ee
This does not work for two reasons.  First, it contains the
contribution to the vacuum functional integral.  Since
\be
k_\mu k'_\nu \vev{A_\mu (k) A_\nu (k')}^\infty = - \xi \delta (k+k')
\ee
we should try instead
\be
\int_k \left[ - \frac{1}{2 \xi^2} k_\mu k_\nu \comp{A_\mu (k) A_\nu
      (-k)} + \frac{1}{2 \xi} \delta (0) \right]
\ee
by subtracting the contribution to the vacuum functional.  (This
amounts to normal ordering.)  Second, the integral over $k$ is UV
divergent.  Subtracting the divergence, we obtain a proper
generalization of (\ref{Oxi-classical}) as
\be\label{Oxi-prime}
\Op'_\xi \equiv \int_k \left[
- \frac{1}{2 \xi^2} k_\mu k_\nu \comp{A_\mu (k) A_\nu (-k)} +
\frac{1}{2 \xi} \delta (0) - \frac{e^2}{2} \frac{f(k/\mu)}{k^4} N_F
\right]
\ee
where $f(k)$ is an arbitrary function that approaches $1$ as $k^2 \to
\infty$, and vanishes for small $k^2$.  $N_F$ is an equation-of-motion
composite operator defined by
\ba\label{NF}
N_F &\equiv& \int_p \K{p} \e^{-\SL} \lb \Tr \left( \comp{\psi (p)}
    \e^{\SL} \right) \Rd{\psi (p)} \right.\nn\\
&&\qquad\qquad\qquad\left.+ \Tr \Ld{\bar{\psi} (-p)} \left(
    \comp{\bar{\psi} (-p)} \e^{\SL} \right)\rb
\ea
This counts the number of fermionic fields:
\ba
&&\vev{N_F\,A_{\mu_1} (k_1) \cdots A_{\mu_L} (k_L) \psi (p_1) \cdots
  \psi (p_N) \bar{\psi} (-q_1) \cdots \bar{\psi} (-q_N)}^\infty\nn\\
&& \qquad\qquad= 2 N \vev{A_{\mu_1} (k_1) \cdots \bar{\psi} (-q_N)}^\infty
\ea

To see the UV finiteness of (\ref{Oxi-prime}), we need the following
equality:
\ba\label{kAkA}
&&- \frac{1}{\xi^2} k_\mu k_\nu \comp{A_\mu (k) A_\nu (-k)} +
\frac{1}{\xi} \delta (0)\nn\\
&&\quad =
\frac{1}{k^4} \left( D(-k) + e \Phi (-k) \right) * \left( D(k) + e \Phi
    (k)\right)
\ea
A proof is given in the first part of \ref{appendix-equality}.

Now, using (\ref{kAkA}), we can rewrite $\Op'_\xi$ as
\be\label{Oxi-prime-alt}
\Op'_\xi = \frac{1}{2} \int_k \frac{1}{k^4} \left[ \left( D(k) + e
        \Phi (k)\right)*\left( D(-k)+e\Phi (-k)\right) - e^2
    f(k/\mu) N_F \right]
\ee
This is manifestly an equation-of-motion composite operator.  To see
the UV finiteness of the integral, we evaluate its correlation
function with elementary fields:
\ba\label{LKh}
&& \vev{\Op'_\xi\,A_{\mu_1} (k_1) \cdots A_{\mu_L} (k_L) \psi
  (p_1) \cdots \psi (p_N) \bar{\psi} (-q_1) \cdots \bar{\psi}
  (-q_N)}^\infty\nn\\ 
&&= - \sum_{i<j} \frac{(k_i)_{\mu_i} (k_i)_{\mu_j}}{k_i^4} \delta
(k_i+k_j) \vev{\cdots \widehat{A_{\mu_i} (k_i)} \cdots
  \widehat{A_{\mu_j} (k_j)} \cdots}^\infty\nn\\
&&\quad + e \sum_{i=1}^L \frac{(k_i)_{\mu_i}}{k_i^4} \sum_{j=1}^N
\lb \vev{\cdots \widehat{A_{\mu_i} (k_i)} \cdots \psi (p_j+k_i)
  \cdots}^\infty \right.\nn\\
&&\,\left. \qquad\qquad\qquad\qquad- \vev{\cdots\widehat{A_{\mu_i}}
      \cdots \bar{\psi} (-q_j+k_i) \cdots}^\infty\rb\\
&&\quad + e^2 \int_k \frac{1}{k^4} \Big[
\sum_{i<j} \lb \vev{\cdots \psi (p_i+k) \cdots \psi
  (p_j-k)\cdots}^\infty\right.\nn\\
&&\qquad\qquad\qquad \left.+ \vev{\cdots\bar{\psi} (-q_i+k) \cdots \bar{\psi} (-q_j -
  k)\cdots}^\infty\rb\nn\\
&&\quad - \sum_{i,j} \vev{\cdots \psi (p_i+k)\cdots\bar{\psi}
  (-q_j-k)\cdots}^\infty + N \left(1 - f(k/\mu)\right)
\vev{\cdots}^\infty \Big]\nn
\ea
Look at the integrand of the $k$-integral without the factor
$\frac{1}{k^4}$.  It vanishes as $k^2 \to \infty$ except for the last
term with the $k$-independent correlation function.  If we choose
$f(\infty) = 1$, its coefficient vanishes, and the $k$-integral
becomes UV finite.  We also note that the integrand (without
$\frac{1}{k^4}$) vanishes at $k=0$ if we choose $f(0) = 0$.  We have
thus justified the counterterm proportional to $N_F$ in
(\ref{Oxi-prime}) and (\ref{Oxi-prime-alt}).

We are left with showing
\be\label{Oxi-Oxi-prime}
\Op_\xi = \Op'_\xi
\ee
This is not valid in general.  The equality requires tuning the $\xi$
dependence of $z_m (\mu)$, $z (\mu)$, and $z_F (\mu)$.  We give the
details in the second part of \ref{appendix-equality}, where we first
note that the difference $\Op_\xi - \Op'_\xi$ is gauge invariant so
that it has three parameters just as the Wilson action.  The
parameters can be taken as $\partial_\xi z_m (\mu), \partial_\xi z
(\mu), \partial_\xi z_F (\mu)$.  Hence, we can satisfy
(\ref{Oxi-Oxi-prime}) by tuning these parameters. For the reader's
convenience, we give the 1-loop results for $z_m (\mu), z (\mu), z_F
(\mu)$ in \ref{qed-1loop}.

Given (\ref{Oxi-corr}) and (\ref{Oxi-Oxi-prime}), the right-hand side
of (\ref{LKh}) gives $- \partial_\xi$ of the correlation function.
This is the renormalized version of the Landau-Khalatnikov relation,
usually given for unrenormalized correlation
functions.\cite{Landau:1955zz}

\section{Dependence of the gauge invariant composite operators on $\xi$ }\label{xi-comp}

In the previous section we have shown that the Wilson action can be
made to satisfy (\ref{Oxi-Oxi-prime}).  We obtain the definition of
$\xi$ independent composite operators by taking an infinitesimal
variation of this equality.

We first recall the definition of $\Op_\xi$ given by (\ref{Oxi}).
Consider $\e^{\SL} (-\Op_\xi)$, and replace $\e^{\SL}$ by $\Op_\Lambda
\e^{\SL}$.  Dividing the result by $\e^{\SL}$, we obtain
\ba\label{dxi}
&&d_\xi \Op_\Lambda 
\equiv \e^{- \SL} \\
&& \times \left[ \partial_\xi + \int_k
    \frac{\K{k}\left(\K{k}-1\right)}{k^4} k_\mu k_\nu \frac{1}{2}
    \frac{\delta^2}{\delta A_\mu (k) \delta A_\nu (-k)} \right] \left(
    \Op_\Lambda \e^{\SL}\right)\nn
\ea
This is a composite operator whose correlation functions with
elementary fields are given by
\be
\vev{d_\xi \Op\,\cdots}^\infty = \partial_\xi
\vev{\Op\, \cdots}^\infty
\ee
Note that we can write 
\be
\Op_\xi = - d_\xi 1
\ee

We next recall the alternative definition (\ref{Oxi-prime-alt}) of
$\Op'_\xi$. Consider $\e^{\SL} (-\Op'_\xi)$ and replace $\e^{\SL}$ by
$\Op_\Lambda \e^{\SL}$.  Dividing the result by $\e^{\SL}$, we obtain
\ba \label{dxiprime}
d'_\xi \Op_\Lambda &\equiv& - \frac{1}{2} \int_k \frac{1}{k^4} \\
&&\times \left[
    \left(D(k) + e \Phi (k)\right)*\left(D(-k)+e\Phi (-k)\right) - e^2
    f(k/\mu) N_F \right] * \Op_\Lambda\nn
\ea
where we define
\ba\label{NFstar}
N_F * \Op_\Lambda &\equiv& \int_p \K{p} \e^{-\SL} \lb
\Tr \left( \comp{\Op \psi (p)} \e^{\SL} \right) \Rd{\psi
  (p)}\right.\nn\\
&&\left.\quad + \Tr \Ld{\bar{\psi} (-p)} \left(\comp{\bar{\psi} (-p)
          \Op} \e^{\SL} \right) \rb
\ea
The latter has the correlation function
\be
\vev{N_F * \Op \cdots}^\infty = 2 N \vev{\Op
  \cdots}^\infty
\ee
where $N$ is the number of $\psi$'s (equivalently $\bar{\psi}$'s)
contained in the dots.  Note that we can write
\be
\Op'_\xi = - d'_\xi 1
\ee
$d'_\xi \Op_\Lambda$ is an
equation-of-motion composite operator with the correlation functions:
\ba\label{dprime-xi-Op}
&& - \vev{d'_\xi \Op \, \cdots} ^\infty\nn\\ 
&&= - \sum_{i<j} \frac{(k_i)_{\mu_i} (k_i)_{\mu_j}}{k_i^4} \delta
(k_i+k_j) \vev{\Op \cdots \widehat{A_{\mu_i} (k_i)} \cdots
  \widehat{A_{\mu_j} (k_j)} \cdots}^\infty\nn\\
&&\, + e \sum_{i=1}^L \frac{(k_i)_{\mu_i}}{k_i^4} \sum_{j=1}^N
\lb \vev{\Op \cdots \widehat{A_{\mu_i} (k_i)} \cdots \psi (p_j+k_i)
  \cdots}^\infty \right.\nn\\
&&\,\left. \qquad\qquad\qquad\qquad- \vev{\Op \cdots\widehat{A_{\mu_i}}
      \cdots \bar{\psi} (-q_j+k_i) \cdots}^\infty\rb\\
&&\, + e^2 \int_k \frac{1}{k^4} \Big[
\sum_{i<j} \lb \vev{\Op \cdots \psi (p_i+k) \cdots \psi
  (p_j-k)\cdots}^\infty\right.\nn\\
&&\qquad\qquad\qquad \left.+ \vev{\Op \cdots\bar{\psi} (-q_i+k) \cdots
      \bar{\psi} (-q_j - k)\cdots}^\infty\rb\nn\\
&&\qquad\qquad\quad - \sum_{i,j} \vev{\Op \cdots \psi (p_i+k)\cdots\bar{\psi}
  (-q_j-k)\cdots}^\infty \nn\\
&&\qquad\qquad\quad + N \left(1 - f(k/\mu)\right)
\vev{\Op \cdots}^\infty \Big]\nn
\ea

We wish to define ``$\xi$-independent'' composite operators such that
$d'_\xi = d_\xi$ acting on them.  With the notation
\be
D_\xi \equiv d_\xi - d'_\xi 
\ee
we then define the ``$\xi$-independence'' of a composite operator
$\Op_\Lambda$ by
\be
D_\xi \Op_\Lambda = 0\label{DxiOp}
\ee
In this notation, (\ref{Oxi-Oxi-prime}) is given by
\be
D_\xi 1 = 0
\ee

In the remainder of this section, we wish to show that we can make any
gauge invariant composite operator $\Op_\Lambda$ ``$\xi$-independent''
by taking an appropriate linear combination with other gauge
invariant composite operators.

Let $\Op_{i \Lambda}\,(i=1,\cdots,n)$ be gauge invariant composite
operators satisfying the WT identity (\ref{WT-qed-op}).  We assume
that this set is closed in the sense that any gauge invariant
composite operator with the same dimension and conserved quantum
numbers can be given as a linear combination of these $n$ composite
operators.  Now, in \ref{WT-Dxi}, we show that $D_\xi \Op_{i \Lambda}$
satisfies the WT identity, if $\Op_{i \Lambda}$ does.  Hence, $D_\xi
\Op_{i \Lambda}$ must be a linear combination of these $n$ gauge
invariant composite operators:
\be
D_\xi \Op_{i \Lambda} = \sum_{j=1}^n C_{ij} (e^2, \xi) \, \Op_{j
  \Lambda}
\ee
where $C_{ij}$ are functions of $e^2$ and $\xi$, independent of the
cutoff $\Lambda$.  Let
\be
\Op'_{i \Lambda} \equiv \sum_{j=1}^n Z_{ij} (e^2, \xi) \Op_{j
  \Lambda}\quad
(i=1,\cdots,n)
\ee
be a new basis of gauge invariant composite operators.  We find
\ba
D_\xi \Op'_{i \Lambda} &=& \sum_{j=1}^n \left(
\partial_\xi Z_{ij} (e^2, \xi)\cdot \Op_{j \Lambda} + Z_{ij} (e^2, \xi)
D_\xi \Op_{j \Lambda}\right)\nn\\
&=& \sum_{j=1}^n \left( \partial_\xi Z_{ij} (e^2, \xi)
+ \sum_k Z_{ik} (e^2,\xi) C_{kj} (e^2,\xi) \right) \Op_{j \Lambda}
\ea
For an arbitrary initial condition $Z_{ij} (e^2, 0)$, we can solve the
homogeneous equations
\be
\partial_\xi Z_{ij} (e^2, \xi)
+ \sum_{k=1}^n Z_{ik} (e^2,\xi) C_{kj} (e^2,\xi) = 0
\ee
Thus, $\Op'_{i\Lambda}$ with these coefficients are the desired
composite operators, that are both gauge invariant and $\xi$
independent.

\section{RG equations for QED\label{RG for QED}}

In this and the next sections we wish to show that the anomalous
dimension of a gauge invariant and $\xi$-independent composite
operator is independent of $\xi$.  Since no discussion of the
$\mu$-dependence of the Wilson action of QED seems available in the
literature, we devote this section for its discussion, rather than
giving it in an appendix.  Then, in the next section, we introduce
anomalous dimensions of composite operators and derive the desired
result.

The Wilson action depends not only on the UV cutoff $\Lambda$, but
also on the renormalization scale $\mu$.  Since $- \mu \partial_\mu
\SL$ can be regarded as an infinitesimal change of $\SL$, it satisfies
the WT identity:
\be \frac{1}{\xi} k_\mu \comp{A_\mu (k) \left(- \mu \partial_\mu
      \SL\right)} = \frac{1}{k^2} \left( D(k) + e \Phi (k) \right) *
\left(- \mu \partial_\mu \SL\right)
\ee
Thus, $- \mu \partial_\mu \SL$ is a gauge invariant composite
operator.

Since the WT identity leaves the Wilson action with three degrees of
freedom, we expect to find three gauge invariant composite operators:
\begin{itemize}
\item[(i)] The composite operator conjugate to the mass parameter $m$ is
    given by
\ba
\Op_m &\equiv& - \e^{-\SL} \Bigg[ \partial_m \e^{\SL} \\
&&\, + \int_p \K{p}
    \left(\K{p}-1\right) \Tr \Ld{\bar{\psi} (-p)} \e^{\SL} \Rd{\psi
      (p)} \frac{-i}{\left(\fey{p} + i m\right)^2} \Bigg]\nn
\ea
This has the following correlation functions:
\be
\vev{\Op_m \cdots}^\infty = - \partial_m \vev{\cdots}^\infty
\ee
where the dots denote a product of elementary fields.
\item[(ii)] The fermion counting operator $N_F$ is given by (\ref{NF}).
\item[(iii)] The third gauge invariant composite operator is given by
\be
N_A - e \Op_e + 2 \xi \Op_\xi
\ee
where $\Op_\xi$, the composite operator conjugate to $\xi$, is defined
by (\ref{Oxi}).  Let us define the other two: $N_A$ is the photon
counting operator
\be\label{NA}
N_A \equiv - \int_k \K{k} \e^{- \SL} \frac{\delta}{\delta A_\mu (k)}
\left( \comp{A_\mu (k)} \e^{\SL} \right)
\ee
and $\Op_e$ is the composite operator conjugate to $e$
\be\label{Oe}
\Op_e \equiv - \partial_e \SL
\ee
$N_A$ and $\Op_e$ have the correlation functions
\ba
\vev{N_A \cdots}^\infty &=& L \vev{\cdots}^\infty\\
\vev{\Op_e \cdots}^\infty &=& - \partial_e \vev{\cdots}^{\infty}
\ea
where $L$ is the number of elementary photon fields in the dots.
\end{itemize}

The WT identities for $\Op_m$ and $N_F$ can be derived most easily
from the WT identity (\ref{WT-qed-corr}) given for the correlation
functions.  Differentiating (\ref{WT-qed-corr}) with respect to $- m$,
we obtain
\ba
&&\frac{1}{\xi} k_\mu \vev{A_\mu (k) \Op_m \cdots}^\infty\nn\\
&& = \frac{1}{k^2} \Big[
\sum_{i=1}^L k_{\mu_i} \delta (k+k_i) \vev{\Op_m \cdots\widehat{A_{\mu_i}
    (k_i)}\cdots}^\infty\\
&&\quad + e \sum_i \lb - \vev{\Op_m \cdots \psi (p_i+k)\cdots}^\infty
+ \vev{\Op_m \cdots\bar{\psi} (-q_i+k)\cdots}^\infty \rb\Big]\nn
\ea
This implies the WT identity for $\Op_m$.  Similarly, we can show the
gauge invariance of $N_F$.  See \ref{combination} for the WT identity
satisfied by the linear combination $N_A - e \Op_e + 2 \xi \Op_\xi$.

Since there are only three composite operators that are gauge
invariant, we must find a linear relation
\be\label{trace}
- \mu \partial_\mu \SL = m \beta_m \Op_m + \gamma_F N_F + \gamma_A
 \left( N_A - e \Op_e + 2 \xi \Op_\xi\right)
\ee
where $\beta_m, \gamma_F, \gamma_A$ are functions of $e^2$ and $\xi$.
This implies nothing but the trace anomaly, giving rise to the usual
RG equation for the correlation functions:
\ba\label{RGeq-qed}
&&\left( - \mu \partial_\mu + \beta \partial_e + \beta_\xi \partial_\xi
    + m \beta_m \partial_m - L \gamma_A - 2 N \gamma_F \right)\nn\\
&& \times \vev{A_{\mu_1} (k_1) \cdots A_{\mu_L} (k_L) \psi (p_1) \cdots \psi
  (p_N) \bar{\psi} (-q_1) \cdots \bar{\psi} (-q_N)}^\infty = 0
\ea
where $\beta_m$ is the anomalous dimension of $m$, and $\gamma_A,
\gamma_F$ are the anomalous dimensions of the photon and fermion
fields.  Note that the beta functions of $e$ and $\xi$ are given in
terms of the anomalous dimension $\gamma_A$ by
\ba
\beta &=& - e \gamma_A\label{betagammaA}\\
\beta_\xi &=& 2 \xi \gamma_A\label{betaxigammaA}
\ea
This implies the RG invariance of the product $e^2 \xi$.

\section{$\xi$ independence of the anomalous dimensions}

In this section, we wish to show that the anomalous dimension of a
composite operator which is both gauge invariant and $\xi$-independent
does not depend on $\xi$.

Let us first introduce the anomalous dimension for an arbitrary
composite operator $\Op_\Lambda$.  We define a new composite operator by
\be\label{dtOp}
d_t \Op_\Lambda \equiv \left(- \mu d_\mu + m \beta_m d_m + \beta d_e
    + \beta_\xi d_\xi - \gamma_F N_F * - \gamma_A N_A * \right) \Op_\Lambda
\ee
where we define
\ba
\e^{\SL} d_\mu \Op_\Lambda &\equiv& \partial_\mu \left( \e^{\SL}
    \Op_\Lambda \right)\\ 
\e^{\SL} d_m \Op_\Lambda &\equiv& \partial_m \left( \e^{\SL}
    \Op_\Lambda \right) + 
\int_p \K{p}\left(\K{p}-1\right) \nn\\
&& \qquad\qquad \times \Tr \Ld{\bar{\psi} (-p)} \left(
    \e^{\SL} \Op_\Lambda \right) 
\Rd{\psi (p)} \frac{- i}{\left(\fey{p} + i m\right)^2} \\
\e^{\SL} d_e \Op_\Lambda &\equiv& \partial_e \left( \e^{\SL} \Op_\Lambda\right)
\ea
so that
\ba
\vev{d_\mu \Op \cdots}^\infty &=& \partial_\mu \vev{\Op
  \cdots}^\infty\\
\vev{d_m \Op \cdots}^\infty &=& \partial_m \vev{\Op \cdots}^\infty\\
\vev{d_e \Op \cdots}^\infty &=& \partial_e \vev{\Op \cdots}^\infty
\ea
Hence, we can write
\be
d_\mu 1 = \partial_\mu \SL,\quad
d_m 1 = - \Op_m,\quad
d_e 1 = - \Op_e
\ee
Using this notation, we can give the trace anomaly (\ref{trace}) as
\be
d_t 1 = 0
\ee

To see that $d_t$ defines the anomalous dimension of a composite
operator, let us suppose $\{ \Op_{i\Lambda} \}_{i=1,\cdots,N}$ are a
basis of composite operators which are both gauge invariant and
$\xi$-independent.  The composite operators mix under the change of
renormalization scale $\mu$:
\be
d_t \Op_{i\Lambda} = \sum_{j=1}^N \gamma_{ij} \Op_{j\Lambda}\label{dtOpi}
\ee
where $\gamma_{ij}$ are functions of $e^2, \xi$.  This implies the RG
equations
\ba
&&\left(- \mu \partial_\mu + \beta \partial_e + \beta_\xi \partial_\xi +
    m \beta_m \partial_m - L \gamma_A - 2 N \gamma_F \right)
\vev{\Op_i\, \cdots}^\infty\nn\\
&&= \sum_j \gamma_{ij} \vev{\Op_j\, \cdots}^\infty
\ea
where the dots stand for products of elementary fields.  

In \ref{independence of dtOp} we show that $d_t \Op_{i\Lambda}$ are
both gauge invariant and $\xi$-independent.  Assuming this result, 
we obtain
\be
\sum_j D_\xi \left( \gamma_{ij} \Op_{j\Lambda} \right) = 
\sum_j \left( \partial_\xi \gamma_{ij} \cdot \Op_{j\Lambda} + \gamma_{ij}
    D_\xi \Op_{j\Lambda} \right) = 0
\ee
Since $D_\xi \Op_{j\Lambda} = 0$, this gives
\be
\partial_\xi \gamma_{ij} = 0
\ee
Thus, the anomalous dimensions are $\xi$-independent.

\section{The axial anomaly}
\label{anomaly}

The axial anomaly is a linear relation among the four gauge invariant
pseudoscalar composite operators:
\[
k_\mu J_{5\mu} (-k),\quad
J_5 (-k),\quad
\comp{\frac{1}{4} F\tilde{F}} (-k),\quad
\Phi_5 (-k)
\]
Except for the last one, which is completely determined by the Wilson
action, the operators must be carefully defined.  Since $e^2 \xi$ is
an RG invariant, specifying the anomalous dimensions of these
composite operators leaves their normalization with arbitrary
dependence on $e^2 \xi$.  Demanding $\xi$-independence, however,
specifies them uniquely only up to constant normalization.  Thus, we
demand these composite operators both gauge invariant and
$\xi$-independent.  (This is automatic for $\Phi_5$.)

\begin{enumerate}
\item \textbf{Axial vector current} $J_{5 \mu} (-k)$ --- The
    asymptotic behavior for large cutoff $\Lambda$ is given by
\ba
J_{5\mu} (-k) &\stackrel{\Lambda \to \infty}{\longrightarrow}& a'_3
(\Lambda) \int_p \bar{\psi} (-p-k) \gamma_5 \gamma_\mu \psi (p)\nn\\
&&\quad + a_5 (\Lambda) \ep_{\mu\alpha\beta\gamma} \int_l A_{\alpha}
(-k-l) l_\beta A_\gamma (l)
\ea
where 
\be
a'_3 (\Lambda) = 1 + \mathrm{O} (e^2),\quad
a_5 (\Lambda) = \mathrm{O} (e^2)
\ee
The tree level value ${a'_3}^{(0)} = 1$ is a normalization condition.
The vanishing of the anomalous dimension specifies the axial vector
current unambiguously.
\item \textbf{Pseudoscalar} $J_5 (-k)$ --- This is determined
    uniquely by the vanishing of the anomalous dimension of
    $m J_5 (-k)$.  The asymptotic behavior is given by
\be
J_5 (-k) \stackrel{\Lambda \to \infty}{\longrightarrow} j(\Lambda)
\int_p \bar{\psi} (-p-k) \gamma_5 \psi (p)
\ee
where $j (\Lambda)$ is normalized as
\be
j (\Lambda) = 1 + \mathrm{O} (e^2)
\ee
\item \textbf{FF dual} $\comp{\frac{1}{4} F\tilde{F}} (-k)$ ---
This is determined uniquely by the vanishing of the anomalous
dimension of $e^2 \comp{\frac{1}{4} F\tilde{F}} (-k)$.  The
asymptotic behavior is given by
\ba
\comp{ \frac{1}{4} F \tilde{F} } (-k)
&\stackrel{\Lambda\to\infty}{\longrightarrow}& f_3 (\Lambda) \int_p
\bar{\psi} (-p-k) \gamma_5 \fey{k} \psi (p)\nn\\
&&\quad + f_5 (\Lambda) k_\mu \ep_{\mu\alpha\beta\gamma} \int_l
A_\alpha (-k-l) l_\beta A_\gamma (l)
\ea
where the normalization condition is\footnote{We choose $\ep_{1234} = 1$.}
\be
f_5 (\Lambda) = 1 + \mathrm{O} (e^2)
\ee
Somewhat unexpectedly, $f_3$ is also of order $1$.  If $f_3$ were of
order $e^2$, the operator would mix with $k_\mu J_{5\mu} (-k)$ under
$d_t$.
\item \textbf{Equation-of-motion} $\Phi_5 (-k)$ is defined by
\ba
\Phi_5 (-k) &\equiv& - \int_p \K{p} \e^{-\SL} \left[ \Tr \left(
        \e^{\SL} \gamma_5 \comp{\psi (p-k)} \right) \Rd{\psi
      (p)}\right.\nn\\
&&\left. \quad + \Tr \Ld{\bar{\psi} (-p)} \left( \e^{\SL}
        \comp{\bar{\psi} (-p-k)} \gamma_5 \right) \right]\\
&\stackrel{\Lambda\to\infty}{\longrightarrow}& \phi'_3 (\Lambda)
\int_p \bar{\psi} (-p-k) \gamma_5 \fey{k} \psi (p)\nn\\
&& + \phi_5 (\Lambda) k_\mu \ep_{\mu\alpha\beta\gamma} \int_l A_\alpha
(-k-l) l_\beta A_\gamma (l)\\
&& + \phi (\Lambda) \int_p \bar{\psi} (-p-k) 2 i m \gamma_5 \psi (p)\nn
\ea
By definition, this is both gauge invariant and $\xi$-independent.  It
has the following correlation functions:
\ba
\vev{\Phi_5 (-k) \cdots}^\infty &=&
- \sum_i \lb \vev{\cdots \gamma_5 \psi (p_i-k) \cdots}^\infty \right.\nn\\
&&\qquad + \left.\vev{\cdots \bar{\psi} (-q_i-k) \gamma_5 \cdots}^\infty \rb
\ea
\end{enumerate}

We define the axial anomaly by
\be
\mathcal{A} (-k) \equiv k_\mu J_{5\mu} (-k) - 2 i m J_5 (-k) - \Phi_5
(-k)
\ee
This is both gauge invariant and $\xi$-independent, and its anomalous
dimension is zero.  Since it vanishes at tree level, it must be
proportional to $e^2$.  The only possibility is a constant multiple of
$e^2 \comp{\frac{1}{4} F \tilde{F}}(-k)$:
\be
\mathcal{A} (-k) = \mathrm{const} \frac{e^2}{(4 \pi)^2}
\comp{\frac{1}{4} F \tilde{F}} (-k)
\ee
This is equivalent to
\ba
a'_3 (\Lambda) - \phi'_3 (\Lambda) &=& \mathrm{const} \frac{e^2}{(4
  \pi)^2} f_3 (\Lambda)\\
a_5 (\Lambda) - \phi_5 (\Lambda) &=& \mathrm{const} \frac{e^2}{(4
  \pi)^2} f_5 (\Lambda)\\
j(\Lambda) + \phi (\Lambda) &=& 0
\ea
The constant is determined by the 1-loop calculation, given in
\ref{anomaly-1loop}, as
\be
\mathrm{const} = - 4
\ee
Hence, we obtain
\be
\mathcal{A} (-k) = - 4 \frac{e^2}{(4 \pi)^2}\comp{ \frac{1}{4} F
  \tilde{F} } (-k)
\ee

\section{Concluding remarks}
\label{conclusion}

We hope we have convinced the reader of the nice compatibility of QED
with ERG.  The main tool provided by ERG is composite operators which
are introduced as infinitesimal variations of the Wilson action.  We
have shown in details how to formulate gauge invariance and gauge
fixing parameter ($\xi$) dependence of composite operators using the
ERG formalism.

Originally we tried to formulate the $\xi$-dependence of QED by
introducing a free scalar field (St\"uckelberg field).  Though it is
straightforward to do this for the unrenormalized QED
\cite{Sonoda:2000kn}, we have not been able to do so for the
renormalized QED.  Once the renormalized Landau-Khalatnikov relation
(\ref{Oxi-Oxi-prime}) is obtained for the Wilson action, it is easy to
guess the $\xi$-independence (\ref{dprime-xi-Op}) for the correlation
functions of physical composite operators.  Expressing the
$\xi$-independence as an operator equation $D_\xi \Op_\Lambda = 0$ has
taken us some efforts.

We expect that most results obtained for QED can be generalized to YM
theories as formulated via ERG.  Equation-of-motion composite
operators will be replaced by BRST exact operators for YM theories.

\ack

Part of the results given here has been presented at
ERG2012 (3--7 September 2012), Aussois, France.  I thank the organizers
for the opportunity.  This work was partially supported by the JSPS
Grants-in-Aid \#22540282 \& \#25400258.

\appendix

\section{ERG in a nutshell}
\label{nutshell}

To make the paper self-contained, we would like to give a short
summary of the ERG formulation applied perturbatively to the $\phi^4$
theory.  See \cite{Igarashi:2009tj} for further details.

\subsection{Wilson action}

The Wilson action is a functional $\SL [\phi]$ of the Fourier
transform $\phi (p)$ of a real scalar field in the four dimensional
Euclidean space.\cite{Wilson:1973jj} We define the correlation
functions by the functional integrals:
\be
\vev{ \cdots }_{\SL} \equiv \int [d\phi]\, \cdots \,\e^{\SL}\Big/\int
[d\phi]\, \e^{\SL}
\ee
The Wilson action splits into the free and interaction parts:
\be
\SL [\phi] = \SF [\phi] + \SI [\phi]
\ee
The free part defined by
\be
\SF [\phi] = - \int_p \frac{p^2 + m^2}{\K{p}} \frac{1}{2} \phi (-p)
\phi (p)
\ee
gives the propagator
\be
\vev{\phi (p) \phi (p')}_{\SF} = \frac{\K{p}}{p^2 + m^2} \cdot \delta
(p+p') 
\ee
(We write $\delta (p)$ for $(2 \pi)^4 \delta^{(4)} (p)$.)  We choose
the cutoff function $\K{p}$ such that it is $1$ for $p^2 < \Lambda^2$,
and it decays fast enough for large momenta so that the correlation
functions are all UV finite.

The dependence of the Wilson action on the UV momentum cutoff
$\Lambda$ is given by the ERG differential equation:\cite{Polchinski:1983gv}
\be\label{phi-ERG}
- \Lambda \frac{\partial}{\partial \Lambda} \e^{\SL}
= \int_p \frac{\D{p}}{p^2 + m^2}  \left[ \frac{p^2+m^2}{\K{p}} \phi (p)
    \frac{\delta}{\delta 
      \phi (p)} + \frac{1}{2} \frac{\delta^2}{\delta \phi (p) \delta
      \phi (-p)} \right] \e^{\SL}
\ee
where
\be
\D{p} \equiv \Lambda \frac{\partial}{\partial \Lambda} \K{p}
\ee
This equation guarantees the $\Lambda$ independence of 
\ba\label{phi-inf}
&&\vev{\phi (p_1) \cdots \phi (p_n)}^\infty \equiv \prod_{i=1}^n
\frac{1}{\K{p_i}} \cdot \Big[ \vev{\phi (p_1) \cdots \phi
  (p_n)}_{\SL}\\
&&\, + \sum_{i<j} \frac{\K{p_i}\left(\K{p_i} - 1\right)}{p_i^2 +
  m^2} \delta (p_i+p_j) \vev{\cdots \widehat{\phi (p_i)} \cdots
  \widehat{\phi (p_j)} \cdots}_{\SL} 
 + \cdots \Big]\nn
\ea
where the dots correspond to terms for which two or more pairs of
$\phi$'s are replaced by
\be\label{phi-mdep-pair}
\frac{\K{p_i}\left(\K{p_i}-1\right)}{p_i^2+m^2} \delta (p_i+p_j)
\ee

\subsection{Parametrization}

To specify a unique solution of (\ref{phi-ERG}), we need to impose an
initial condition at a particular $\Lambda$.  Alternatively, we can
specify the asymptotic behavior of $\SL$ for large
$\Lambda$:\footnote{The coefficients are expanded in powers of
  $m^2/\Lambda^2$ and momenta divided by $\Lambda$.}
\ba
\SL &\stackrel{\Lambda \to \infty}{\longrightarrow}& - \int_p \, (p^2
+ m^2) \frac{1}{2} \phi (p) \phi (-p)\nn\\
&& + \int_p \left( a_2 (\Lambda) + z(\Lambda) p^2 \right) \frac{1}{2}
\phi (p) \phi (-p)\nn\\
&& + \int_{p_1,p_2,p_3} a_4 (\Lambda) \frac{1}{4!} \phi (p_1) \phi
(p_2) \phi (p_3) \phi (-p_1-p_2-p_3)
\ea
The existence of a perturbative solution of this type, characterized
by the vanishing of higher dimensional terms, amounts to perturbative
renormalizability of the theory.\cite{Sonoda:2002pb}

The $\Lambda$ dependence of $a_2, z, a_4$ is determined by
(\ref{phi-ERG}).  Their values at an arbitrary scale $\mu$ parametrize
the Wilson action:
\[
a_2 (\mu),\quad z (\mu),\quad a_4 (\mu)
\]
If we choose a convention
\be
a_2 (\mu) = z (\mu) = 0,\quad a_4 (\mu) = - \lambda
\ee
we obtain
\be
\lb\begin{array}{c@{~=~}l}
a_2 (\Lambda) & \frac{\lambda}{4} (\Lambda^2 - \mu^2) \int_p
\frac{\Delta (p)}{p^2} - m^2 \frac{\lambda}{(4\pi)^2} \ln \frac{\Lambda}{\mu}\\
z (\Lambda) & 0\\
a_4 (\Lambda) & - \lambda \left( 1 + 3 \frac{\lambda}{(4 \pi)^2} \ln
    \frac{\Lambda}{\mu} \right)
\end{array}
\right.
\ee
up to 1-loop.

\subsection{Composite operators\label{app-phi4-comp}}

Composite operators $\Op_\Lambda [\phi]$ are $\Lambda$-dependent
functionals that can be interpreted as infinitesimal variations of the
Wilson action.  Hence, its $\Lambda$ dependence is given by
\ba
- \Lambda \frac{\partial}{\partial \Lambda} \left(\Op_\Lambda
    \e^{\SL}\right) 
&=& \int_p \frac{\D{p}}{p^2 + m^2}  \\
&&\times \left[ \frac{p^2+m^2}{\K{p}} \phi (p)
    \frac{\delta}{\delta 
      \phi (p)} + \frac{1}{2} \frac{\delta^2}{\delta \phi (p) \delta
      \phi (-p)} \right] \left( \Op_\Lambda \e^{\SL}\right)\nn
\ea
which is obtained from (\ref{phi-ERG}) by changing $\SL$
infinitesimally by $\Op_\Lambda$.  Corresponding to (\ref{phi-inf}),
the cutoff independent correlation functions are given by
\ba
&&\vev{\Op\,\phi (p_1) \cdots \phi (p_n)}^\infty \equiv \prod_{i=1}^n
\frac{1}{\K{p_i}} \cdot \Big[ \vev{\Op_\Lambda \phi (p_1) \cdots \phi
  (p_n)}_{\SL}\\
&&\, + \sum_{i<j} \frac{\K{p_i}\left(\K{p_i} - 1\right)}{p_i^2 +
  m^2} \delta (p_i+p_j) \vev{\Op_\Lambda \cdots \widehat{\phi (p_i)} \cdots
  \widehat{\phi (p_j)} \cdots}_{\SL} 
 + \cdots \Big]\nn
\ea

The simplest example of a composite operator is
\begin{equation}
\comp{\phi (p)} \equiv \frac{1}{\K{p}} \phi (p) +
\frac{1-\K{p}}{p^2 + m^2} \frac{\delta \SL}{\delta \phi (-p)}
\end{equation}
This gives the same correlation functions as the elementary field
$\phi (p)$:
\be
\vev{[\phi (p)] \phi (p_1) \cdots \phi (p_n)}^\infty
= \vev{\phi (p) \phi (p_1) \cdots \phi (p_n)}^\infty
\ee

Given a composite operator $\Op_\Lambda$, its product with an
elementary field $\phi (p)$ is given by the composite operator
\ba
\comp{\Op\,\phi (p)} &\equiv& \Op_\Lambda\,\comp{\phi (p)} + \frac{1-\K{p}}{p^2
  + m^2} \frac{\delta}{\delta \phi (-p)} \Op_\Lambda\\
&=& \e^{-\SL} \left( \frac{1}{\K{p}} \phi (p) + \frac{1-\K{p}}{p^2 +
      m^2} \frac{\delta}{\delta \phi (-p)} \right) \left( \Op_\Lambda
    \,\e^{\SL} \right)\nn
\ea
As the second expression shows, $\e^{\SL} \comp{\Op \phi (p)}$ is
obtained from $\e^{\SL} \comp{\phi (p)}$ by changing $\SL$
infinitesimally by $\Op_\Lambda$.  We find
\be
\vev{\left[\Op \phi (p)\right] \phi (p_1) \cdots \phi (p_n)}^\infty
= \vev{\Op\,\phi (p) \phi (p_1) \cdots \phi (p_n)}^\infty
\ee

Given an arbitrary composite operator $\Op_\Lambda (p)$ dependent on a
momentum $p$, we can define a corresponding equation-of-motion
composite operator by
\be\label{EoM-operator}
\Op'_\Lambda \equiv - \int_p \K{p} \e^{-\SL} \frac{\delta}{\delta \phi
  (p)} \left( \e^{\SL} \Op_\Lambda (p) \right)
\ee
In the correlation function with elementary fields, $\Op'_\Lambda$
replaces each $\phi (p_i)$ by $\Op_\Lambda (p_i)$:
\be
\vev{\Op'\,\phi (p_1) \cdots \phi (p_n)}^\infty
= \sum_{i=1}^n \vev{\phi (p_1) \cdots \Op (p_i)
  \cdots \phi (p_n)}^\infty
\ee
We can rewrite the definition of the composite operator $\comp{\phi
  (p)}$ as the operator equation
\be
\left(p^2 + m^2\right) \comp{\phi (p)} - \frac{\delta \SI}{\delta \phi
  (-p)} = - \K{p} \frac{\delta \SL}{\delta \phi (-p)}
\ee
where the right-hand side is an equation-of-motion composite operator
(\ref{EoM-operator}) with $\Op_\Lambda (q) = \delta (p-q)$.

\subsection{Trace anomaly --- beta functions and anomalous dimension}

The logarithmic derivative $- \mu \partial_\mu \SL$ is a composite
operator corresponding to the trace anomaly.\cite{Sonoda:2006ai}
Since it is a dimension $4$ operator, we can expand it as
\begin{equation}\label{phi-trace}
- \mu \partial_\mu \SL = \beta (\lambda) \Op_\lambda + \beta_m
(\lambda, m^2) \Op_m + \gamma (\lambda) \N
\end{equation}
where
\begin{eqnarray}
\Op_\lambda &\equiv& - \e^{-\SL} \partial_\lambda \e^{\SL}\\
\Op_m &\equiv& - \e^{-\SL} \left[ \partial_{m^2} - \int_p
    \frac{K\left(K-1\right)}{(p^2+m^2)^2} \frac{1}{2}
    \frac{\delta^2}{\delta \phi (p)\delta \phi (-p)} \right]
\e^{\SL}\label{phi-Om}\\
\N &\equiv& - \int_p \K{p} \e^{-\SL} \frac{\delta}{\delta \phi (p)}
\left( \e^{\SL} [\phi (p)] \right)
\end{eqnarray}
are the three independent composite operators with dimension up to
$4$.  The second term of (\ref{phi-Om}) is necessary because of the
$m^2$ dependence of (\ref{phi-mdep-pair}).  The $\Lambda$ independent
correlation functions of these operators are given by
\ba
\vev{\Op_\lambda\,\phi (p_1) \cdots \phi (p_n)}^\infty &=&
- \partial_\lambda \vev{\phi (p_1) \cdots \phi (p_n)}^\infty\\
\vev{\Op_m\,\phi (p_1) \cdots \phi (p_n)}^\infty &=&
- \partial_{m^2} \vev{\phi (p_1) \cdots \phi (p_n)}^\infty\\
\vev{\N\,\phi (p_1) \cdots \phi (p_n)}^\infty &=&
n \vev{\phi (p_1) \cdots \phi (p_n)}^\infty
\ea
Thus, (\ref{phi-trace}) gives
\be
\left( - \mu \partial_\mu + \beta (\lambda) \partial_\lambda + \beta_m
    (\lambda,m^2) \partial_{m^2} - n \gamma (\lambda) \right)
\vev{\phi (p_1) \cdots \phi (p_n)}^\infty = 0
\ee
which is a standard RG equation, where $\beta$ and $\beta_m$ are the
beta functions of $\lambda$ and $m^2$, and $\gamma$ is the anomalous
dimension of $\phi$.

Up to 1-loop, we obtain\footnote[1]{Note that $\beta_m$ is not
  proportional to $m^2$.  For mass independence, we must require $a_2
  (\Lambda)$ not to have terms proportional to $\mu^2$.}
\be
\lb
\begin{array}{c@{~=~}l}
\beta_m & - m^2 \frac{\lambda}{(4 \pi)^2} + \frac{\lambda}{2} \mu^2
\int_p \frac{\Delta (p)}{p^2}\\
\beta & - 3 \frac{\lambda^2}{(4 \pi)^2}\\
\gamma & 0
\end{array}\right.
\ee

\section{Summary of notations for composite operators}
\label{appendix-notation}

Given a composite operator $\Op_\Lambda$, there are three ways of
generating more composite operators.

\subsection{Products with elementary fields}

\ba
\comp{A_\mu (k) \Op} &\equiv& \e^{-\SL} \left( \frac{A_\mu (k)}{\K{k}} 
    + \frac{1 - \K{k}}{k^2} \right.\nn\\
&&\qquad \left. \times\left( \delta_{\mu\nu} - (1-\xi)
        \frac{k_\mu k_\nu}{k^2} \right) \frac{\delta}{\delta A_\nu
      (-k)} \right) \left( \e^{\SL} \Op_\Lambda \right)\\
\comp{\psi (p) \Op} &\equiv& \e^{-\SL} \left( \frac{\psi (p)}{\K{p}} 
    + \frac{1-\K{p}}{\fey{p} + i m} \Ld{\bar{\psi} (-p)}
\right) \left( \e^{\SL} \Op_\Lambda \right)\\
\comp{\bar{\psi} (-p) \Op} &\equiv& \left( \Op_\Lambda
    \e^{\SL} \right) \left( \frac{\bar{\psi} (-p)}{\K{p}}  + \Rd{\psi
      (p)} \frac{1 - \K{p}}{\fey{p} + i m} \right) \e^{-\SL}
\ea

\subsection{Derivatives}

\ba
d_\mu \Op_\Lambda &\equiv& \e^{- \SL} \partial_\mu \left( \e^{\SL} \Op_\Lambda
\right)\\
d_e \Op_\Lambda &\equiv& \e^{- \SL} \partial_e \left( \e^{\SL}
    \Op_\Lambda \right)\\ 
d_m \Op_\Lambda &\equiv& \e^{- \SL} \left[ \partial_m \left( \e^{\SL}
        \Op_\Lambda 
    \right)  + \int_p \K{p} \left( \K{p}-1\right) \right.\nn\\
&&\quad\qquad \times \left.\Tr \Ld{\bar{\psi} (-p)}
\left( \e^{\SL} \Op_\Lambda \right) \Rd{\psi (p)} \frac{-i}{\left(
      \fey{p} + i m \right)^2} \right]\\ 
d_\xi \Op_\Lambda &\equiv& \e^{-\SL} \left[ \partial_\xi + \int_k
    \K{k} \left(\K{k}-1\right) \frac{k_\mu k_\nu}{k^4} \right.\nn\\
&&\quad\qquad \times \left.\frac{1}{2}
    \frac{\delta^2}{\delta A_\mu (k) \delta A_\nu (-k)} \right] \left(
    \e^{\SL} \Op_\Lambda\right)
\ea

\subsection{Equations-of-motion}

\ba
D(k) * \Op_\Lambda &\equiv& \e^{-\SL} \K{k}  k_\mu (-) \frac{\delta}{\delta
  A_\mu (-k)} \left( \e^{\SL} \Op_\Lambda \right)\\
N_A (k) * \Op_\Lambda &\equiv& \e^{-\SL} \int_l \K{l} (-) \frac{\delta}{\delta
  A_\mu (l)} \left( \e^{\SL} \comp{A_\mu (k+l) \Op} \right)\\
N_\psi (k) * \Op_\Lambda &\equiv& \e^{-\SL} \int_p \K{p} \Tr \left(
    \e^{\SL} \comp{\psi  (p+k) \Op} \right) \Rd{\psi (p)}\\
N_{\bar{\psi}} (k) * \Op_\Lambda &\equiv& \e^{-\SL} \int_p \K{p} \Tr
\Ld{\bar{\psi} (-p)} \left( \e^{\SL} \comp{ \Op \bar{\psi} (-p+k)}
\right)\\
\Phi (k) *  \Op_\Lambda &\equiv& \left(- N_\psi (k) + N_{\bar{\psi}}
    (k) \right) * \Op_\Lambda\\ 
N_F (k) * \Op_\Lambda &\equiv& \left( N_\psi (k) + N_{\bar{\psi}} (k)
\right) * 
\Op_\Lambda\\ 
d'_\xi \Op_\Lambda &\equiv& - \frac{1}{2} \int_k \frac{1}{k^4} \Big[
    \left(D(k) +  e \Phi (k)\right) * \left( D(-k) + e \Phi (-k) \right)
\nn\\
&&\qquad\qquad - f(k/\mu) e^2 N_F \Big] * \Op_\Lambda
\ea

\subsection{Commutation relations}

Derivatives and equation-of-motion $N_A (k)*$, $N_\psi(k') *$,
$N_{\bar{\psi}} (k'') *$ commute among themselves.  The only exception
is $D(k)*$ which commutes with everything except for $N_A (k')*$.  We
find in particular
\be
\label{DNA-commutator}
D(k) * N_A (0) - N_A (0) * D(k) = - D(k)
\ee

\section{Proof of $\Op_\xi = \Op'_\xi$}
\label{appendix-equality}

\subsection{Proof of (\ref{kAkA})}

Using the WT identity (\ref{WT-qed}), we obtain
\be
- \frac{1}{\xi^2} k_\mu k_\nu \comp{A_\mu (k) A_\nu (-k)} =
\frac{1}{\xi} \frac{1}{k^2} k_\mu \comp{A_\mu (k) \left( D(-k) + e
      \Phi (-k) \right)}
\ee
Using the definitions (\ref{Phi}, \ref{D}) and (\ref{Dstar},
\ref{Phistar}), we obtain
\ba
\comp{A_\mu (k) D(-k)} &=& D(-k) * \comp{A_\mu (k)} - k_\mu \delta
(0)\\
\comp{A_\mu (k) \Phi (-k)} &=& \Phi (-k) * \comp{A_\mu (k)}
\ea
Hence, we obtain
\ba
&&- \frac{1}{\xi^2} k_\mu k_\nu \comp{A_\mu (k) A_\nu (-k)} \nn\\
&&\quad= \frac{1}{k^2} \left(D(-k) + e \Phi (-k)
\right) * \frac{1}{\xi} k_\mu \comp{A_\mu (k)} - \frac{1}{\xi} \delta (0)\nn\\
&&\quad= \frac{1}{k^4} \left(D(-k) + e \Phi (-k) \right) * \left( D(k) + e
    \Phi (k) \right) - \frac{1}{\xi} \delta (0)
\ea
where we have used (\ref{WT-qed}) once more.  This is the desired equality.

\subsection{Perturbative determination of $\partial_\xi
  z_m, \partial_\xi z, \partial_\xi z_F$}

We wish to show that we can make $\Op_\xi = \Op'_\xi$ by choosing the
$\xi$ dependence of the parameters $z_m (\mu), z (\mu), z_F (\mu)$
appropriately.  Though this appendix augments the discussion in
section \ref{xi}, we use the notations $d_\xi$ (\ref{dxi}) and
$d'_\xi$ (\ref{dxiprime}) introduced in section \ref{xi-comp}.
Moreover, we use the result of \ref{WT-Dxi}.  We ask the reader to
glance over the beginning of section \ref{xi-comp} for $d_\xi,
d'_\xi$, and bare with us for our assuming the result of \ref{WT-Dxi}.

We first show that the difference $\Op_\xi - \Op'_\xi$ satisfies the
WT identity:
\be\label{WT-difference}
\frac{1}{\xi} k_\mu \comp{A_\mu (k) \left( \Op_\xi - \Op'_\xi \right)}
= \frac{1}{k^2} \left( D(k) + e \Phi (k) \right) * \left( \Op_\xi -
    \Op'_\xi \right) 
\ee
Since
\be
\Op_\xi = - d_\xi 1,\quad \Op'_\xi = - d'_\xi 1
\ee
we find
\be
\Op_\xi - \Op'_\xi = - D_\xi 1
\ee
In \ref{WT-Dxi} we show the gauge invariance of $D_\xi \Op_\Lambda$
for any gauge invariant composite operator $\Op_\Lambda$.  Since
$\Op_\Lambda = 1$ is gauge invariant, $\Op_\xi - \Op'_\xi$ is gauge
invariant, satisfying (\ref{WT-difference}).

Now, the gauge invariant $\Op_\xi - \Op'_\xi$ has three degrees of
freedom, just as the Wilson action itself.  Hence, we can make it
vanish by tuning three parameters.  To see this more explicitly, we
examine the asymptotic behavior of $\Op_\xi - \Op'_\xi$ for $\Lambda$
much bigger than the momenta of the fields.  Let $\Op^{(n)}_\Lambda$
be the $n$-loop part of $\Op_\Lambda$.  We then obtain, for $n \ge 1$,
\be
\Op^{(n)}_\xi \stackrel{\Lambda \to \infty}{\longrightarrow}
- \partial_\xi S_\Lambda^{(n)} + \int_k
\frac{\K{k}\left(\K{k}-1\right)}{k^4} k_\mu k_\nu \frac{1}{2}
\frac{\delta^2 \SL^{(n-1)}}{\delta A_\mu (k) \delta A_\nu (-k)}
\ee
and
\ba
\Op'^{(n)}_\xi &\stackrel{\Lambda \to \infty}{\longrightarrow}&
- \int_k \lb \left( \frac{1-\K{k}}{k^2} \right)^2 k_\mu k_\nu
\frac{1}{2} \frac{\delta^2 \SL^{(n-1)}}{\delta A_\mu (k) \delta A_\nu
  (-k)} \nn\right.\\
&&\left. \qquad\qquad+ \frac{e^2}{2} \frac{f(k/\mu)}{k^4} N_F^{(n-1)} \rb
\ea
where the integration over $k$ provides one extra loop.  Note
$\partial_\xi S_\Lambda^{(n)}$ is parametrized by the three parameters
$\partial_\xi z_m^{(n)} (\mu), \partial_\xi z^{(n)}
(\mu), \partial_\xi z_F^{(n)} (\mu)$.  Thus, by tuning these, we can
make $\Op_\xi - \Op'_\xi$ vanish at $n$-loop.  Since $\Op_\xi =
\Op'_\xi$ at tree level, we have proven $\Op_\xi = \Op'_\xi$ by
mathematical induction on the number of loops.

We give the 1-loop results of tuning in \ref{qed-1loop}.

\section{Gauge invariance of $D_\xi \Op_\Lambda$}
\label{WT-Dxi}

The result of this appendix is necessary for the previous appendix and
section \ref{xi-comp}.  We wish to show that $D_\xi \Op_\Lambda \equiv
(d_\xi - d'_\xi) \Op_\Lambda$, where  $\Op_\Lambda$ satisfies the WT
identity (\ref{WT-qed-op}), also satisfies the WT identity:
\be\label{WT-DxiOp}
\frac{1}{\xi} k_\mu \comp{ A_\mu (k) D_\xi \Op } =
\frac{1}{k^2} \left( D(k) + e \Phi (k) \right) * D_\xi \Op_\Lambda
\ee

In the following we will adopt a formal algebraic approach.
Alternatively, we could derive (\ref{WT-DxiOp}) by examining the
correlation functions of $D_\xi \Op_\Lambda$ with elementary fields.
Using the definition of $D_\xi$ and the WT identity satisfied by
$\Op_\Lambda$, we can compute
\[
\frac{1}{\xi} k_\mu \vev{ A_\mu (k) D_\xi \Op\, \cdots}^\infty
\]
and show that this has the expected form as the WT identity for $D_\xi
\Op_\Lambda$. 

Let us instead give a formal algebraic proof.  First, let
$\Op_\Lambda$ be an arbitrary composite operator, not necessarily
gauge invariant.  By definition we have
\be
d_\xi \left( k_\mu \comp{ A_\mu (k) \Op} \right)
= k_\mu \comp{A_\mu (k) d_\xi \Op}
\ee
Hence, we obtain
\be\label{result1}
\frac{1}{\xi} k_\mu \comp{A_\mu (k) d_\xi \Op} - d_\xi \left(
    \frac{1}{\xi} k_\mu \comp{A_\mu (k) \Op} \right) = \frac{1}{\xi^2}
k_\mu \comp{A_\mu (k) \Op}
\ee

From (\ref{Phi}, \ref{D}) and (\ref{Dstar}, \ref{Phistar}), we obtain
\ba
k_\mu \comp{A_\mu (k) D(l) * \Op} &=& D(l)* k_\mu \comp{A_\mu (k) \Op}
- k^2 \delta (k+l) \Op_\Lambda\\
k_\mu \comp{A_\mu (k) \Phi (l) * \Op} &=& \Phi (l) * k_\mu \comp{A_\mu
  (k) \Op}
\ea
Hence, we obtain
\ba\label{result3}
&&k_\mu \comp{A_\mu (k) \left(D(l) + e \Phi (l)\right)*\left(D(-l)+e
      \Phi (-l)\right)* \Op}\nn\\
&&= \left( D(l) + e \Phi (l)\right)* k_\mu \comp{A_\mu (k)\left(D(-l)+e
      \Phi (-l)\right)* \Op}\nn\\
&&\quad - k^2 \delta(k+l) \left(D(k) + e \Phi
    (k)\right)*\Op_\Lambda\nn\\
&&=  \left( D(l) + e \Phi (l)\right)* \lb
\left( D(-l) + e \Phi (-l)\right) * k_\mu \comp{A_\mu (k) \Op} - k^2
\delta (k-l) \Op_\Lambda \rb\nn\\
&&\quad - k^2 \delta(k+l) \left(D(k) + e \Phi
    (k)\right)*\Op_\Lambda\nn\\
&&= \left( D(l) + e \Phi (l)\right)* 
\left( D(-l) + e \Phi (-l)\right) * k_\mu \comp{A_\mu (k) \Op}\nn\\
&&\quad - k^2 \left( \delta (k+l) + \delta (k-l) \right) \left(D(k) + e \Phi
    (k)\right)*\Op_\Lambda
\ea
From (\ref{NFstar}), we also obtain
\be\label{result4}
k_\mu \comp{A_\mu (k) N_F * \Op} = N_F * k_\mu \comp{A_\mu (k) \Op}
\ee
Recalling the definition of $d'_\xi \Op_\Lambda$ (\ref{dxiprime}):
\ba
d'_\xi \Op_\Lambda &\equiv& \frac{1}{2} \int_k \frac{1}{k^4} \left[
\left(D(k) + e \Phi (k)\right)*\left(D(-k) + e \Phi
    (-k)\right)\nn\right.\\
&&\qquad\qquad\left. - e^2
f(k/\mu) N_F \right] * \Op_\Lambda
\ea
we obtain, from (\ref{result3}) and (\ref{result4}),
\be\label{result2}
k_\mu \comp{A_\mu (k) d'_\xi \Op} - d'_\xi k_\mu \comp{A_\mu (k) \Op}
= \frac{1}{k^2} \left( D(k) + e \Phi (k) \right) * \Op_\Lambda
\ee

Thus, from (\ref{result1}) and (\ref{result2}), we obtain
\ba
&&\frac{1}{\xi} k_\mu \comp{A_\mu (k) D_\xi \Op} - 
D_\xi \left(\frac{1}{\xi} k_\mu \comp{A_\mu (k) \Op}\right)\nn\\
&& = \frac{1}{\xi} \lb
\frac{1}{\xi} k_\mu \comp{A_\mu (k) \Op} - \frac{1}{k^2} \left( D(k) +
    e \Phi (k) \right) * \Op_\Lambda \rb
\ea
This is valid for any composite operator $\Op_\Lambda$.  If
$\Op_\Lambda$ satisfies the WT identity, the right-hand side vanishes,
and we obtain
\be
\frac{1}{\xi} k_\mu \comp{A_\mu (k) D_\xi \Op} = D_\xi \left(
    \frac{1}{k^2} \left( D(k) + e \Phi (k)\right) * \Op_\Lambda
\right)
\ee
Since $D_\xi$ commutes with $D(k) + e \Phi (k)$\footnote{$d_\xi$
  commutes with $D(k) *$ and $\Phi (k) *$.  $d'_\xi$, consisting of 
  $D(k') *$, $\Phi (k') *$, and $N_F *$, also commutes with $D(k)*$
  and $\Phi(k)*$.}, we
obtain
\be
\frac{1}{\xi} k_\mu \comp{A_\mu (k) D_\xi \Op} = \frac{1}{k^2} \left(
    D(k) + e \Phi (k) \right)* D_\xi \Op_\Lambda 
\ee
Thus, $D_\xi \Op_\Lambda$ satisfies the WT identity.

\section{Gauge invariance of $N_A - e \Op_e + 2 \xi \Op_\xi$}
\label{combination}

In this appendix we prove the gauge invariance of the linear
combination $N_A - e \Op_e + 2 \xi \Op_\xi$, where $N_A$ is defined by
(\ref{NA}), $\Op_e$ by (\ref{Oe}), and $\Op_\xi$ by (\ref{Oxi}).  It
is the simplest if we derive the WT identity (\ref{WT-qed-op-corr})
for the correlation functions for $\Op = N_A - e \Op_e + 2 \xi
\Op_\xi$.

From the original WT identity (\ref{WT-qed-corr}), we first obtain
\ba
&&\frac{1}{\xi} k_\mu \vev{A_\mu (k) N_A \cdots}^\infty\nn\\
&&= \frac{1}{k^2} \Big[
\sum_i k_{\mu_i} \delta (k+k_i) \vev{\left(N_A + 2\right) \cdots
  \widehat{A_{\mu_i} (k_i)} \cdots}^\infty\nn\\
&&\qquad + e \sum_i \lb - \vev{\left(N_A + 1\right) \cdots \psi (p_i +
  k)\cdots}^\infty \right.\\
&&\left.\qquad\qquad + \vev{\left(N_A + 1\right) \cdots \bar{\psi} (-q_i
  + k) \cdots}^\infty \rb\nn \Big]
\ea
Differentiating (\ref{WT-qed-corr}) by $e$ and multiplying by $e$, we
obtain
\ba
&&\frac{1}{\xi} k_\mu \vev{A_\mu (k) \left(-e \Op_e\right) \cdots}^\infty\nn\\
&&= \frac{1}{k^2} \Big[ \sum_i k_{\mu_i} \delta (k+k_i) \vev{\left(-e
      \Op_e\right) 
  \cdots \widehat{A_{\mu_i} (k_i)} \cdots}^\infty\\
&&\qquad + e \sum_i \lb - \vev{\left(- e \Op_e +1\right) \cdots \psi
  (p_i+k)\cdots}^\infty \right.\nn\\
&&\qquad\qquad \left.+ \vev{\left(- e \Op_e + 1\right) \cdots \bar{\psi}
  (-q_i+k)\cdots}^\infty \rb\nn
\ea
Finally, differentiating (\ref{WT-qed-corr}) by $- \xi$ and
multiplying by $2 \xi$, we obtain
\ba
&&\frac{1}{\xi} k_\mu \vev{A_\mu (k) \left( 2 \xi \Op_\xi + 2 \right)
  \cdots}^\infty\nn\\
&&= \frac{1}{k^2} \Big[ \sum_i k_{\mu_i} \delta (k+k_i) \vev{2 \xi
  \Op_\xi \cdots \widehat{A_{\mu_i} (k_i)} \cdots}^\infty\\
&&\quad + e \sum_i \lb - \vev{2 \xi \Op_\xi \cdots \psi
  (p_i+k)\cdots}^\infty + \vev{2 \xi \Op_\xi \cdots \bar{\psi}
  (-q_i+k)\cdots}^\infty\rb \Big]\nn
\ea
Summing the three equations together, we obtain the WT identity for
the linear combination:
\ba &&\frac{1}{\xi} k_\mu \vev{A_\mu (k) \left( N_A - e \Op_e + 2 \xi
      \Op_\xi\right) \cdots}^\infty\nn\\
&&= \frac{1}{k^2} \Big[ \sum_i k_{\mu_i} \delta (k+k_i) \vev{\left(
      N_A - e \Op_e + 2 \xi \Op_\xi\right) \cdots \widehat{A_{\mu_i}
    (k_i)} \cdots}^\infty\nn\\
&&\quad + e \sum_i \lb - \vev{\left( N_A - e \Op_e + 2 \xi
      \Op_\xi\right) \cdots \psi (p_i+k)\cdots}^\infty\right.\\
&&\qquad \left. + \vev{\left( N_A - e \Op_e + 2 \xi
      \Op_\xi\right) \cdots \bar{\psi} (-q_i+k) \cdots}^\infty \rb
\Big]\nn
\ea

\section{Asymptotic coefficients at 1-loop}
\label{qed-1loop}

At 1-loop, the WT identity (\ref{WT-qed}) alone gives the following
results:
\ba a_2^{(1)} (\Lambda) &=& \frac{e^2}{(4 \pi)^2} \left[ - 2 (4 \pi)^2
    \Lambda^2 \int_p \frac{\Delta
      (p) \left(1 - K(p)\right)}{p^2} + 2 m^2 \right]\\
\tilde{z}^{(1)} (\Lambda) &=& - z^{(1)} (\Lambda) +\frac{2}{3}
\frac{ e^2 }{(4 \pi)^2}\\
\frac{1}{e} a_3^{(1)} (\Lambda) &=& - z_F^{(1)} (\Lambda) \nn\\
&& \, - 
\frac{e^2}{(4\pi)^2} \left( \xi
    (4 \pi)^2 \int_p \frac{K(p) \left(1 - K(p)\right)^2}{p^4} + \frac{3 -
      \xi}{4} \right)\\
\frac{1}{e^2} a_4^{(1)} (\Lambda) &=& \frac{4}{3} \frac{e^2}{(4 \pi)^2} \ea
where
\ba
z^{(1)} (\Lambda) &=& \frac{8}{3} \frac{e^2}{(4 \pi)^2} \ln
\frac{\Lambda}{\mu} + z^{(1)}\\
z_F^{(1)} (\Lambda) &=& 2 \xi \frac{e^2}{(4 \pi)^2} \ln
\frac{\Lambda}{\mu} + z_F^{(1)} \\
z_m^{(1)} (\Lambda) &=& 2 (3+\xi) \frac{e^2}{(4 \pi)^2} \ln
\frac{\Lambda}{\mu} + z_m^{(1)} 
\ea
Imposing $\Op_\xi = \Op'_\xi$ (\ref{Oxi-Oxi-prime}) further, we obtain
\ba
z^{(1)} &=& \frac{e^2}{(4 \pi)^2} \mathcal{Z}\\
z_F^{(1)} &=& \frac{e^2}{(4 \pi)^2} \left[ \mathcal{Z}_F + \xi \lb
    \frac{1}{4} - (4\pi)^2 \int_k \frac{1}{k^4} \left( \left(1 -
            K(k)\right)^2 - f(k) \right) \rb \right]\\
z_m^{(1)} &=& \frac{e^2}{(4 \pi)^2} \left[  \mathcal{Z}_m - \xi (4\pi)^2 \int_k
    \frac{1}{k^4} \left( \left(1 - K(k)\right)^2 - f(k) \right) \right] \ea
where the numerical constants $\mathcal{Z}, \mathcal{Z}_F,
\mathcal{Z}_m$ are still left arbitrary.

\section{Beta functions and anomalous dimensions}
\label{beta}

The $\mu$-dependence of the Wilson action is given by (\ref{trace})
\be\label{appendix-trace}
- \mu \partial_\mu \SL = m \beta_m \Op_m + \gamma_F N_F + \gamma_A
 \left( N_A - e \Op_e + 2 \xi \Op_\xi\right)
\ee

\subsection{Anomalous dimensions at 1-loop}
\label{anomalous dimensions at 1-loop}

To extract $\beta_m, \gamma_F, \gamma_A$, we compare the
asymptotic behaviors of both hand sides.  At tree level, we find
\ba
m \Op_m^{(0)} &\stackrel{\Lambda \to \infty}{\longrightarrow}& i m
\int d^4 x\, \bar{\psi} \psi\\
N_A^{(0)} - e \Op_e^{(0)} + 2 \xi \Op_\xi^{(0)} &\stackrel{\Lambda \to
  \infty}{\longrightarrow}& \frac{1}{2} \int d^4 x\, F_{\mu\nu}
F_{\mu\nu}\\
N_F^{(0)} &\stackrel{\Lambda \to \infty}{\longrightarrow}& 2 \int d^4
x\, \bar{\psi} \left( \frac{1}{i} \fey{\partial} + i m - e \fey{A}
\right) \psi
\ea
On the other hand, at 1-loop level, the asymptotic coefficients of the
previous appendix give
\ba
&&- \mu \partial_\mu \SL^{(1)} \stackrel{\Lambda \to
  \infty}{\longrightarrow} \frac{e^2}{(4 \pi)^2} \int d^4 x\, \Bigg[\nn\\
&&\quad
\frac{4}{3} \cdot \frac{1}{2} F_{\mu\nu} F_{\mu\nu} + 2 \xi \bar{\psi}
\left( \frac{1}{i} \fey{\partial} + i m - e \fey{A}
\right) \psi + 6 i m \bar{\psi} \psi \Bigg]
\ea
Hence, we obtain the following 1-loop results:
\ba
\beta_m^{(1)} &=& 6 \frac{e^2}{(4 \pi)^2}\\
\gamma_A^{(1)} &=& \frac{4}{3} \frac{e^2}{(4 \pi)^2}\\
\gamma_F^{(1)} &=& \xi \frac{e^2}{(4 \pi)^2}
\ea

\subsection{Anomalous dimensions of composite operators}

The trace anomaly (\ref{appendix-trace}) implies
\ba
&&\left( - \mu \partial_\mu + \beta \partial_e + \beta_\xi \partial_\xi
    + m \beta_m \partial_m - L \gamma_A - 2 N \gamma_F \right)\\
&& \times \vev{A_{\mu_1} (k_1) \cdots A_{\mu_L} (k_L) \psi (p_1) \cdots \psi
  (p_N) \bar{\psi} (-q_1) \cdots \bar{\psi} (-q_N)}^\infty = 0\nn
\ea
for the correlation functions.  Differentiating these with respect to
the parameters $m, e, \xi$, we can extract the anomalous dimensions
for the conjugate operators:
\ba
d_t \Op_m &=& - \beta_m \Op_m\\
d_t \Op_e &=& - (\partial_e \beta_m) m \Op_m + \partial_e (e \gamma_A)
\Op_e - \partial_e (2 \xi \gamma_A) \Op_\xi \nn\\
&& - (\partial_e \gamma_A) N_A - (\partial_e \gamma_F) N_F\\
d_t \Op_\xi &=& - (\partial_\xi \beta_m) m \Op_m + \partial_\xi (e
\gamma_A) \Op_e - \partial_\xi (2 \xi \gamma_A) \Op_\xi \nn\\
&& -
(\partial_\xi \gamma_A) N_A - (\partial_\xi \gamma_F) N_F
\label{dtOxi}
\ea
Since the number of elementary fields does not change under an
infinitesimal change of $\mu$, we obtain
\be
d_t N_A = d_t N_F = 0
\ee

\subsection{$\xi$ dependence of $\beta_m, \gamma_A, \gamma_F$}

We recall that the gauge invariance of $- \mu d_\mu 1 = -
\mu \partial_\mu \SL$ implies (\ref{betagammaA}) and
(\ref{betaxigammaA}).  We can obtain the $\xi$-dependence of the
anomalous dimensions $\beta_m, \gamma_A, \gamma_F$ from the
$\xi$-independence of the action $\Op_\xi = \Op'_\xi$.  $d_t \Op_\xi$
is given by (\ref{dtOxi}).  To compute $d_t \Op'_\xi$, we recall
\be
\Op'_\xi \equiv \int_k \left[ - \frac{1}{2 \xi^2} k_\mu k_\nu
    \comp{A_\mu (k) A_\nu (-k)} + \frac{1}{2 \xi} \delta (0) -
    \frac{e^2}{2} \frac{f (k/\mu)}{k^4} N_F \right]
\ee
Since
\ba
d_t \comp{A_\mu (k) A_\nu (-k)} &=& 2 \gamma_A  \comp{A_\mu (k) A_\nu
  (-k)}\\
d_t \frac{1}{\xi} &=& - 2 \gamma_A \frac{1}{\xi}\\
d_t e^2 &=& - 2 \gamma_A e^2\\
d_t N_F &=& 0
\ea
we obtain
\be
d_t \Op'_\xi = - 2 \gamma_A \Op'_\xi + \frac{e^2}{2} \int_k 
\frac{1}{k^4} \mu \partial_\mu f(k/\mu)\, N_F
\ee
To calculate the integral, we only need the asymptotic values $f(0) =
0, f(\infty) = 1$.  We find
\be\label{f-integral}
\int_k \frac{1}{k^4} \mu \partial_\mu f(k/\mu) = - \frac{2}{(4 \pi)^2}
\ee
Hence, we obtain
\be
d_t \Op'_\xi = - 2 \gamma_A \Op'_\xi - \frac{e^2}{(4 \pi)^2} N_F
\ee
Comparing this with (\ref{dtOxi}), we obtain
\be\label{beta-xi-dependence}
\partial_\xi \beta_m = \partial_\xi \gamma_A = 0,\quad
\partial_\xi \gamma_F = \frac{e^2}{(4 \pi)^2}
\ee

\section{Gauge invariance and $\xi$-independence of $d_t \Op_\Lambda$}
\label{independence of dtOp}

Let us assume that $\Op_\Lambda$ is both gauge invariant and
$\xi$-independent, satisfying (\ref{WT-qed-op}) and (\ref{DxiOp})
\ba
\frac{1}{\xi} \comp{k_\mu A_\mu (k) \Op} &=& \frac{1}{k^2}
\left( D(k) + e \Phi (k) \right) * \Op_\Lambda\label{appendix-WTOp}\\
D_\xi \Op_\Lambda &=& 0
\ea
We wish to show first that $d_t \Op$ satisfies the WT identity, and
then that $d_t \Op$ is $\xi$-independent.

\subsection{WT identity for $d_t \Op$}

$d_t \Op$ is defined by (\ref{dtOp}):
\be
d_t \Op \equiv \left( - \mu d_\mu + m \beta_m d_m + \beta d_e +
    \beta_\xi d_\xi - \gamma_F N_F * - \gamma_A N_A * \right) \Op
\ee
We find
\ba
&&\frac{1}{\xi} k_\mu \comp{A_\mu (k) d_t \Op}\nn\\
&&= \frac{1}{\xi} k_\mu \comp{A_\mu (k) \left( - \mu d_\mu + \beta d_e
      + \beta_\xi d_\xi + \beta_m d_m - \gamma_A N_A* - \gamma_F N_F*
  \right) \Op}\nn\\
&&= \lb - \mu d_\mu + \beta d_e
      + \beta_\xi \left(d_\xi + \frac{1}{\xi}\right) + \beta_m d_m \rb
\frac{1}{\xi} k_\mu \comp{A_\mu (k) \Op}\nn\\
&&\quad - \gamma_A \frac{1}{\xi} k_\mu \comp{A_\mu (k) N_A* \Op} -
\gamma_F \frac{1}{\xi} k_\mu \comp{A_\mu (k) N_F*\Op}
\ea
Since
\ba
\frac{1}{\xi} k_\mu \comp{A_\mu (k) N_A * \Op} &=& N_A* \frac{1}{\xi}
k_\mu \comp{A_\mu (k) \Op} + \frac{1}{\xi} k_\mu \comp{A_\mu (k)
  \Op}\\
\frac{1}{\xi} k_\mu \comp{A_\mu (k) N_F * \Op} &=& N_F* \frac{1}{\xi}
k_\mu \comp{A_\mu (k) \Op} 
\ea
we obtain
\ba
&&\frac{1}{\xi} k_\mu \comp{A_\mu (k) d_t \Op}
= \lb  - \mu d_\mu + \beta d_e
      + \beta_\xi \left(d_\xi + \frac{1}{\xi}\right)\right.\nn\\
&&\quad + \beta_m d_m 
- \gamma_A N_A * - \gamma_A - \gamma_F N_F * \Bigg\rbrace \frac{1}{\xi} k_\mu
\comp{A_\mu (k) \Op}
\ea
Using the WT identity (\ref{appendix-WTOp}) and the commutator
(\ref{DNA-commutator}), we obtain
\ba
\frac{1}{\xi} k_\mu \comp{A_\mu (k) d_t \Op}
&=& \frac{1}{k^2} \left( D(k) + e \Phi (k) \right) *
\Bigg\lbrace
 - \mu d_\mu + \beta d_e + \beta_\xi \left(d_\xi +
    \frac{1}{\xi}\right)\nn\\
&&  + \beta_m m d_m - \gamma_A N_A * - \gamma_A -
\gamma_F N_F * \Bigg\rbrace \Op\nn\\
&&+ \frac{1}{k^2} \left( - \gamma_A D(k)  + \beta \Phi (k) 
\right) * \Op
\ea
Finally, using (\ref{betagammaA}, \ref{betaxigammaA}), we obtain
\ba
\frac{1}{\xi} k_\mu \comp{A_\mu (k) d_t \Op} &=& \frac{1}{k^2} \left(
    D(k) + e \Phi (k) \right) * \left( d_t \Op + \gamma_A \Op
\right)\nn\\
&& - \frac{1}{k^2} \gamma_A \left(  D(k) + e \Phi (k) \right) *
\Op\nn\\
&=& \frac{1}{k^2} \left( D(k) + e \Phi (k) \right) * d_t \Op
\ea
Thus, $d_t \Op_\Lambda$ satisfies the WT identity.

\subsection{$\xi$-independence of $d_t \Op_\Lambda$}

We will prove
\be\label{Dxidt}
D_\xi d_t \Op_\Lambda = 0
\ee
by deriving the commutator
\be\label{dtDxi-commutator}
d_t D_\xi - D_\xi d_t = - 2 \gamma_A D_\xi
\ee
Assuming $D_\xi \Op_\Lambda = 0$, this commutator gives immediately
(\ref{Dxidt}).

We recall the definition
\be
d_t \equiv - \mu d_\mu - e \gamma_A d_e + 2 \xi \gamma_A d_\xi +
\beta_m m d_m - \gamma_A N_A * - \gamma_F N_F *
\ee
where the $\xi$-dependence of $\gamma_A, \beta_m, \gamma_F$ is given
by (\ref{beta-xi-dependence}). Since $d_\xi$ commutes with derivatives
and $N_A *, N_F*$, we obtain
\be\label{dxidt-comm}
d_\xi d_t - d_t d_\xi = 2 \gamma_A d_\xi - (\partial_\xi \gamma_F) N_F
*  = 2 \gamma_A d_\xi - \frac{e^2}{(4 \pi)^2} N_F *
\ee

To compute the commutator of $d_t$ and $d'_\xi$, we recall
\ba
d'_\xi &\equiv& - \frac{1}{2} \int_k \frac{1}{k^4} \left[ \left(D(k) + e
        \Phi (k)\right) * \left( D(-k) + e \Phi (-k) \right) *
\right.\nn\\
&&\qquad\qquad \left.- f(k/\mu) e^3 N_F * \right]
\ea
As a preparation, we compute the commutator
\ba
&& \left( D(-k) + e \Phi (-k) \right) * d_t - d_t \left( D(-k) + e
    \Phi (-k) \right) *\nn\\
&&= e \gamma_A \Phi (-k) - \gamma_A \left( D(-k) * N_A * - N_A * D(-k)
    * \right)
\ea
Using the commutator (\ref{DNA-commutator}), we get
\ba
&&\left( D(-k) + e \Phi (-k) \right) * d_t - d_t \left( D(-k) + e
    \Phi (-k) \right) * \nn\\
&& = \gamma_A \left( D(-k) + e \Phi (-k) \right) *
\ea
Thus, we obtain
\ba
&& \left( D(k) + e \Phi (k) \right) * \left( D(-k) + e \Phi (-k)
\right) * d_t\nn\\
&&= \left( D(k) + e \Phi (k) \right) * \left[
d_t \left( D(-k) + e \Phi (-k)\right) * 
+ \gamma_A \left( D(-k) + e \Phi (-k)\right) * \right]\nn\\
&&= d_t \left( D(k) + e \Phi (k) \right) * \left( D(-k) + e \Phi (-k)
\right) * \nn\\
&&\quad + 2 \gamma_A \left( D(k) + e \Phi (k) \right) * \left( D(-k) +
    e \Phi (-k) \right) * 
\label{dt-commutator1}
\ea
Since $N_F *$ commutes with $d_t$, we obtain
\ba
&& f(k/\mu) e^2 N_F * d_t\nn\\
&&= d_t \lb f(k/\mu) e^2 N_F *\rb + \left( \mu \partial_\mu f(k/\mu)
+ 2 \gamma_A \right) e^2 N_F * 
\label{dt-commutator2}
\ea

Hence, using (\ref{dt-commutator1}, \ref{dt-commutator2}), we obtain
\ba\label{dprimexidt-comm}
d'_\xi d_t &=& d_t d'_\xi - \frac{1}{2} \int_k \frac{1}{k^4} \Bigg[ 2
\gamma_A \left(D(k) + e \Phi (k)\right) * \left( D(-k) + e \Phi (-k)
\right) * \nn\\
&& \quad - 2 \gamma_A f(k/\mu) e^2 N_F * - \mu \left(\partial_\mu
    f(k/\mu)\right) e^2 N_F * \Bigg]\nn\\
&=& d_t d'_\xi + 2 \gamma_A d'_\xi + \frac{1}{2} e^2 \int_k
\frac{1}{k^4} \mu \left( \partial_\mu f(k/\mu) \right) N_F *\nn\\
&=& d_t d'_\xi + 2 \gamma_A d'_\xi - \frac{e^2}{(4 \pi)^2} N_F *
\ea
where we have used (\ref{f-integral}).

Combining (\ref{dxidt-comm}) and (\ref{dprimexidt-comm}), we obtain
\be
d_t \left( d_\xi - d'_\xi \right) - \left( d_\xi - d'_\xi \right) d_t
= - 2 \gamma_A \left( d_\xi - d'_\xi \right)
\ee
This is the desired commutator (\ref{dtDxi-commutator}).

\section{Axial anomaly at 1-loop}
\label{anomaly-1loop}

Up to 1-loop, we obtain the following results:
\begin{itemize}
\item[(i)] $J_{5 \mu}$
\ba
a'_3 (\Lambda) &=& 1 + \frac{e^2}{(4 \pi)^2} \left[ - 2 \xi \ln
    \frac{\Lambda}{\mu} + A \right.\nn\\
&&\quad\left. + \xi (4 \pi)^2 \int_k \frac{1}{k^4} \lb
    \left(1-K(k)\right)^3 - f(k) \rb \right]\\
a_5 (\Lambda) &=& \frac{e^2}{(4\pi)^2} \frac{-8}{3}
\ea
$a_5$ is determined by $a'_3$ by gauge invariance.  The numerical
constant $A$ cannot be fixed by imposing the vanishing of the 1-loop
anomalous dimension.
\item[(ii)] $J_5$
\ba
j (\Lambda) &=& 1 + \frac{e^2}{(4 \pi)^2} \left[ - 2 (3 + \xi) \ln
    \frac{\Lambda}{\mu} + B\right.\nn\\
&&\quad \left. + \xi (4 \pi)^2 \int_k \frac{1}{k^4} \lb
    \left(1 - K(k)\right)^3 - f(k) \rb \right]
\ea
where the numerical constant $B$ cannot be fixed by demanding the
vanishing of the anomalous dimension of $m J_5$ at 1-loop.
\item[(iii)] $\comp{\frac{1}{4} F \tilde{F}}$
\ba
f_3 (\Lambda) &=& - \frac{9}{8} + \frac{e^2}{(4 \pi)^2} \left[
\left( - 3 + \frac{9}{4} \xi \right) \ln \frac{\Lambda}{\mu} \right.\nn\\
&&\quad \left.-
\frac{9}{8} \xi (4 \pi)^2 \int_k \frac{1}{k^4} \lb
    \left(1 - K(k)\right)^3 - f(k) \rb + C \right]\\
f_5 (\Lambda) &=& 1 + \frac{e^2}{(4 \pi)^2} D
\ea
where $C, D$ cannot be fixed by demanding the vanishing of the anomalous
dimension of $e^2 \comp{\frac{1}{4} F\tilde{F}}$ at 1-loop.
\item[(iv)] $\Phi_5$ --- this is unambiguously determined by the Wilson
    action.
    \ba \phi'_3 (\Lambda) &=& 1 + \frac{e^2}{(4 \pi)^2} \left[ - 2 \xi
        \ln
        \frac{\Lambda}{\mu} - \frac{3}{4} - \mathcal{Z}_F \right.\nn\\
    &&\qquad\qquad \left.+ \xi (4\pi)^2
        \int_k \frac{1}{k^4} \lb (1-K)^3 - f\rb \right]\\
    \phi_5 (\Lambda) &=& \frac{e^2}{(4 \pi)^2} \frac{4}{3}\\
    \phi (\Lambda) &=& - 1 + \frac{e^2}{(4 \pi)^2} \left[ 2 (3+\xi)
        \ln \frac{\Lambda}{\mu} + \mathcal{Z}_m \right.\\
&& \left.\, - \xi (4 \pi)^2 \int_k
        \frac{1}{k^4} \lb (1-K)^3 - f \rb 
     + 3 (4 \pi)^2 \int_k \frac{K(1-K)^2}{k^4}
    \right] \nn
\ea 
    where the undetermined numerical constants $\mathcal{Z}_F,
    \mathcal{Z}_m$ are carried over from \ref{qed-1loop}.
\end{itemize}

We thus obtain, up to 1-loop,
\ba
a'_3 (\Lambda) - \phi'_3 (\Lambda) &=& \frac{e^2}{(4 \pi)^2} \left( A
    + \frac{3}{4} + \mathcal{Z}_F \right)\\
a_5 (\Lambda) - \phi_5 (\Lambda) &=& \frac{e^2}{(4 \pi)^2} (-4)\\
j(\Lambda) + \phi (\Lambda) &=& \frac{e^2}{(4 \pi)^2} \left[ B +
    \mathcal{Z}_m + 3 (4\pi)^2 \int_k \frac{K(1-K)^2}{k^4} \right]
\ea
The last equation must vanish, and we obtain
\be
B = \mathcal{Z}_m - 3 (4 \pi)^2 \int_k \frac{K(1-K)^2}{k^4}
\ee
Comparing the first two equations with
\ba
\mathrm{const} \frac{e^2}{(4 \pi)^2} f_3^{(0)} (\Lambda) &=&
\frac{e^2}{(4 \pi)^2} \,\mathrm{const} \,\frac{-9}{8} \\
\mathrm{const} \frac{e^2}{(4 \pi)^2} f_5^{(0)} (\Lambda) &=&
\frac{e^2}{(4 \pi)^2} \,\mathrm{const} 
\ea
we obtain
\be
\mathrm{const} = - 4
\ee
and
\be
A = \frac{15}{4} - \mathcal{Z}_F
\ee


\section*{References}

\begin{thebibliography}{00}
\bibitem{Igarashi:2009tj}
  Igarashi Y, Itoh K and Sonoda H
  2010 \textit{Prog.\ Theor.\ Phys.\ Suppl.}  {\bf 181} 1
  [arXiv:0909.0327 [hep-th]]
\bibitem{Rosten:2010vm}
  Rosten O J
  2012 \textit{Phys.\ Rept.}  {\bf 511} 177
  [arXiv:1003.1366 [hep-th]]
\bibitem{Collins}
  Collins J 1984 \textit{Renormalization} (Cambridge University Press)
\bibitem{Wilson:1973jj}
  Wilson K G and Kogut J B
  1974 \textit{Phys.\ Rept.}  {\bf 12} 75
\bibitem{Polchinski:1983gv}
  Polchinski J
  1984 \textit{Nucl.\ Phys.} B {\bf 231} 269
\bibitem{Morris:2006in}
  Morris T R and Rosten O J
  2006 \textit{J.\ Phys.} A {\bf 39} 11657
  [hep-th/0606189]
\bibitem{Becchi:1996an}
  Becchi C
  ``On the construction of renormalized gauge theories using
  renormalization group techniques'' 
  [hep-th/9607188]
\bibitem{Ellwanger:1994iz}
  Ellwanger U
  1994 \textit{Phys.\ Lett.} B {\bf 335} 364
  [hep-th/9402077]
\bibitem{Bonini:1993kt}
  Bonini M, D'Attanasio M and Marchesini G
  1994 \textit{Nucl.\ Phys.} B {\bf 418} 81
  [hep-th/9307174]
\bibitem{Reuter:1994sg}
  Reuter M and Wetterich C
  1994 \textit{Nucl.\ Phys.} B {\bf 427} 291
\bibitem{Sonoda:2007dj}
  Sonoda H
  2007 \textit{J.\ Phys.} A {\bf 40} 9675
  [hep-th/0703167]
\bibitem{Adler:1969er}
  Adler S L and Bardeen W A
  1969 \textit{Phys.\ Rev.}  {\bf 182} 1517
\bibitem{Zee:1972zt}
  Zee A
  1972 \textit{Phys.\ Rev.\ Lett.}  {\bf 29} 1198
\bibitem{Landau:1955zz}
  Landau L D and Khalatnikov I M
  1956 \textit{Sov.\ Phys.\ JETP} {\bf 2} 69
   [1955 \textit{Zh.\ Eksp.\ Teor.\ Fiz.}  {\bf 29} 89]
\bibitem{Svidzinskii}
  Svidzinskii A V 1955 Thesis, L'vov State University
\bibitem{Bogoliubov}
 Bogoliubov N N and Shirkov D V 1959 \textit{Introduction to the Theory
   of Quantized Fields} (Wiley-Interscience) Chapter VII,
 Section 40
\bibitem{Zumino:1959wt}
  Zumino B
  1960 \textit{J.\ Math.\ Phys.}  {\bf 1} 1
\bibitem{Sonoda:2006ai}
  Sonoda H
  2007 \textit{J.\ Phys.} A {\bf 40} 5733
  [hep-th/0612294]
\bibitem{Sonoda:2002pb}
  Sonoda H
  2003 \textit{Phys.\ Rev.} D {\bf 67} 065011
  [hep-th/0212302]
\bibitem{Sonoda:2000kn}
  Sonoda H
  2001 \textit{Phys.\ Lett.} B {\bf 499} 253
  [hep-th/0008158]
\end{thebibliography}
\end{document}